\documentclass[journal,draftcls,onecolumn,12pt,twoside]{IEEEtranTCOM}
\normalsize
\usepackage{graphicx}
\usepackage{verbatim}
\usepackage{upgreek}
\usepackage{amssymb,amsmath}
\usepackage{color}
\usepackage{setspace}
\usepackage{epstopdf}
\usepackage{bm}
\usepackage{cite}
\usepackage{array,color}
\usepackage{algorithm}
\usepackage{algpseudocode}
\usepackage{amsmath}
\usepackage{graphics}
\usepackage{epsfig}
\usepackage{ulem}
\usepackage{subfigure}
\begin{document}
\title{Belief Propagation based Joint Detection and Decoding for Resistive Random Access Memories}

\author{Ce Sun, Kui Cai, Guanghui Song, Tony Q. S. Quek, Zesong Fei
\thanks{ Ce Sun, Zesong Fei are with the School of Information and Electronics, Beijing Institute of Technology, Beijing 10081, China. E-mail:\{3120160396, feizesong\}@bit.edu.cn.}
\thanks{Kui Cai, Guanghui Song are with Science and Math Cluster, Singapore University of Technology and Design, Singapore. E-mail:\{gsong2017@gmail.com, cai\_kui@sutd.edu.sg\}}
\thanks{Tony Q. S. Quek is with Dept. of Information Systems Technology and Design, Singapore University of Technology and Design, Singapore. E-mail:tonyquek@sutd.edu.sg}
\thanks{This paper was supported by Singapore Ministry of Education Academic Research Fund Tier 2 MOE2019-T2-2-123 and RIE2020 Advanced Manufacturing and Engineering (AME) programmatic grant A18A6b0057, and also support by the National Natural Science Foundation of China under Grant No.61871032 .}
}% <-this % stops a space

\markboth{IEEE Transactions on Communications}%
{Submitted paper}
\maketitle

\begin{abstract}
  Despite the great promises that the resistive random access memory (ReRAM) has shown as the next generation of non-volatile memory technology, its crossbar array structure leads to a severe sneak path interference to the signal read back from the memory cell. In this paper, we first propose a novel belief propagation (BP) based detector for the sneak path interference in ReRAM. Based on the conditions for a sneak path to occur and the dependence of the states of the memory cells that are involved in the sneak path, a Tanner graph for the ReRAM channel is constructed, inside which specific messages are updated iteratively to get a better estimation of the sneak path affected cells. We further combine the graph of the designed BP detector with that of the BP decoder of the polar codes to form a joint detector and decoder. Tailored for the joint detector and decoder over the ReRAM channel, effective polar codes are   constructed using the genetic algorithm. Simulation results show that the BP detector can effectively detect the cells affected by the sneak path, and the proposed polar codes and the joint detector and decoder can significantly improve the error rate performance of ReRAM.

% Based on the dependence among cells caused by the sneak-path interference, a Tanner graph for the ReRAM channel is constructed, inside which specific messages are updated iteratively to get a better estimate of the sneak-path affected cells.
  \vspace{2ex}
\end{abstract}
\section{Introduction}
Resistive random access memory (ReRAM) is an emerging nonvolatile memory (NVM) technique that demonstrates great potential to enhance the data storage density and reduce the power consumption and read/write speed of the memory \cite{1.ReRAM}. In ReRAM, the binary input user information bits of `0's and `1's are represented by the high and low resistance states of the memory cell, respectively. A key feature of ReRAM is the simple crossbar structure, which offers huge density gain. However, the isolation among memory cells in the crossbar array is poor, which results in the sneak path problem \cite{2.Sneak_path}. In particular, when a high resistance state (logic 0) cell is read, an alternative current passing through other cells in the memory array may distort the measurement, leading to an erroneous decision of a low resistance state (logic 1) instead. Unlike the typical noises/inteference of data storage systems, the sneak path interference is a unique problem to be addressed as it is data-dependent. More precisely, the occurrence of the sneak path depends on the input user data. A most widely adopted approach to tackle the sneak path problem is to introduce the cell selectors, which is an electrical device that allows current to go through only in one direction. However, the cell selectors tend to fail randomly due to the imperfections in the memory fabrication and maintenance processes. As a result, the sneak path interference may still occur and corrupt the signals read back from the memory array.

%In the literatures, the sneak-path problem was addressed at various system layers. For example, alternative memory architectures were proposed, which included a modification of the cell technology and/or the entire array structure\cite{3.4} \cite{3.5}. Other approaches concentrated on low-level electric analysis to clean distorted measurements \cite{6.6}-\cite{6.9}.

% Error correction coding and signal processing can provide cost-effective solutions to tackle different noises and interference of data storage systems.

Channel coding and signal processing serve as a cost-effective means to mitigate the various noise and interference in data storage systems. In the literatures, information theoretical framework to mitigate the sneak path problem for the crossbar resistive memories arrays was studied in \cite{11.theo_ReRAM}, which calculated the error probability as a function of the array dimensions and the distribution of the input data bits. Authors in \cite{13.sf1} formulated a probabilistic sneak path model with the assumption that the cell selector fails with a certain probability, and developed several channel detection schemes for the proposed channel model. However, both the sneak path channel model and the subsequent detectors did not capture and utilize exactly the dependence of the states of the memory cells that are involved in the sneak path. In \cite{14.sf2}, adaptive thresholding schemes were proposed to adaptively change the detection threshold using side information provided by certain pilot cells, which helps to determine the occurrence/nonoccurrence of the sneak path. The drawback of introducing pilot cells is that it will incur additional redundancies. In \cite{17.code1}, based on an estimation of the channel sneak path rate ({\it i.e.} the ratio of the high resistance cells that are affected by the sneak path interference), an elementary signal estimator (ESE), which is essentially a soft-output channel detector, is proposed which can calculate the loglikelihood ratio (LLR) of each channel bit. However, the ESE still works on the basis of treating the sneak path interference as an independent and identically distributed (\textit{i.i.d.}) noise.

Error correction codes (ECCs) have been widely applied to correct the memory cell errors. However, so far limited work has been done on error correction coding for ReRAM. In particular, a non-stationary polar coding scheme for ReRAM was proposed by \cite{18.code2}, which is mainly aimed to address the raw bit error rate (BER) diversity among different bitlines, rather than tacking the sneak path interference. In \cite{17.code1}, an across-array coding strategy is proposed, which assigns a codeword to multiple memory arrays. In this way, the sneak path interference induced dependence of the states of the memory cells within an ECC codeword can be reduced. However, the drawback of the across-array coding strategy is an unavoidable increase of the read/write latency.

The belief propagation (BP) algorithm \cite{6.BP} has been extensively studied for the decoding of channel codes, such as the turbo codes and low-density parity-check (LDPC) codes \cite{19.LDPC}. The BP algorithm updates the current marking state of the entire Markov Random Field (MRF) by using the mutual information between nodes and nodes. It is an approximate calculation based on MRF. The algorithm is an iterative method, which can solve the problem of probabilistic inference of the probabilistic graph model, and the propagation of all information can be realized in parallel. After multiple iterations, the reliability of all nodes no longer changes, and it is said that the mark of each node at this time is the optimal mark, and the MRF has also reached a state of convergence. For ReRAM systm, we observe that there are data dependence in the storage array because of sneak path. And the storage array can be represented in the form of a matrix. Data dependence and matrix representation are similar to the check matrix of LDPC. Inspired by LDPC decoder, we consider that construct a Tanner graph for ReRAM affected by sneak path for detection. In this work, by exploring the conditions for the sneak path to occur in the crossbar array and the dependence of the states of the memory cells that are involved in the sneak path, we propose a novel BP detector to detect the memory cells that are affected by the sneak path interference. The proposed BP based detection consists of three steps. First, we detect memory cells in high resistance states that are not affected by the sneak path interference. After that, we detect memory cells that are definitely in the low resistance states (instead of being converted into the low resistance state by the sneak path interference) based on the conditions for the sneak path to occur. In the third step, we define the remaining memory cells as the selector failure nodes (SFNs) and the sneak path nodes (SPNs), and use them as two sets of nodes in the Tanner graph. They are connected by specific edges based on the conditions required to generate the sneak paths. A third set of node named the detection-aiding node (DAN) is further added to the Tanner graph and connected with the SFNs. Messages about the probability of the faulty selectors and sneak path affected cells are passed among the SFNs, SPNs and DANs in the Tanner graph and being updated iteratively to get a better estimation of the sneak path affected cells.

To further improve the reliability of ReRAM, we cascade the graph of the designed BP detector with that of the BP decoder to form a joint detection and decoding scheme. Messages are passed in the combined graph iteratively, with the BP detector mainly mitigating the sneak path interference and the BP decoder correcting the errors caused by channel additive noise. Next, as an example, we employ the polar codes as ECCs for the ReRAM channel. Conventionally, polar codes are designed using density evolution (DE) or polarization weight (PW) approximation for the successive cancellation (SC) decoding algorithm over the symmetric channels. However, the ReRAM channel is asymmetry by nature. In this work, we adopt the genetic algorithm (GenA) to construct polar codes for the BP based joint detector and decoder over the ReRAM channel.

The rest of this paper is organized as follows. In Section II,  we introduce the system model of ReRAM with the sneak path interference. In Section III, we present the proposed BP detector.  The BP based joint detection and decoding scheme is further proposed in Section IV, and polar codes tailored for the ReRAM channel are also constructed in this section. The performance of the proposed BP based joint detection and decoding with the designed polar codes is evaluated in Section V. Finally, Section VI concludes the paper.

\section{System Model}

In this work, we consider an $M\times N$ resistive crossbar array. The cell $(i,j)$, denoted by $U_{i,j}$, lies at the intersection of row $i$ and column $j$, $1\leq i\leq M, 1\leq j\leq N$. Define matrix $\textbf{X} \in {\left\{ {0,1} \right\}^{M \times N}}$ to be the input user data array, with its entry $x_{i,j}$ being stored in $U_{i,j}$. The cell $U_{i,j}$ is in the high resistance state $R_0$ when $x_{i,j}=0$ while it is in the low resistance state $R_1$ when $x_{i,j}=1$. We consider the input data distribution as \textit{i.i.d.} Bernoulli ($q$), {\it i.e.}, $\Pr \left( {{x_{i,j}} = 1} \right) = q$ and $\Pr \left( {{x_{i,j}} = 0} \right) = 1 - q$.

\begin{figure}
  \centering
  \includegraphics[width=2.1in, height=2in]{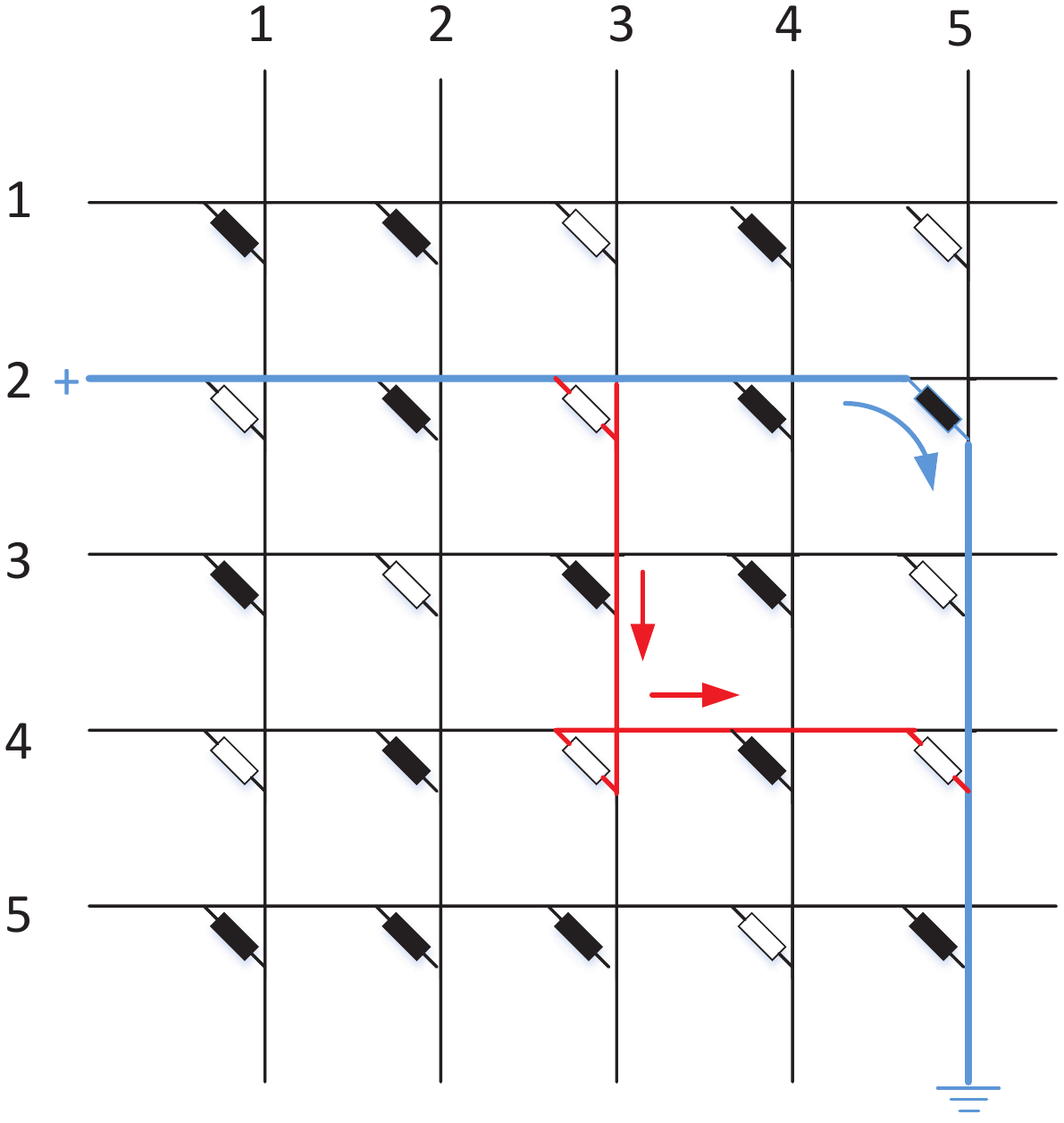}
  \caption{Example of a $5\times5$ memristor array, where (2,5) is a selected cell for reading, and the blue line shows the desired path for resistance measuring. However, $(2,3)\rightarrow (4,3)\rightarrow (4,5)$ forms a sneak path (red line) in parallel to the desired path.}\label{fig_sneak_path}
\end{figure}

During the read operation, the resistance of the target memory cell $U_{i,j}$ is measured by applying voltage upon the cell. If the target cell is in the high resistance state, there may occur parallel paths traversing through unselected cells with low resistance states. This will cause the the target cell to be erroneously identified as in the low resistance state, thus leading to the sneak path problem. An example is illustrated in Fig.\ref{fig_sneak_path}, where the high resistance is marked with black and the low resistance is in white. The cell (2,5) with high resistance is the target cell for reading and the blue line shows the desired path for resistance measuring. However,  $(2,3)\rightarrow (4,3)\rightarrow (4,5)$ forms a sneak path (red line) in parallel of the desired path. Due to the low resistance cells in the parallel path, the measurement of resistance of the cell $(2,5)$ will be corrupted.

In order to tackle the sneak path problem, a most popular method is to add a cell selector in series to each memory cell. As the cell selector only allows the current to flow in one direction, it can help to avoid the occurrence of the sneak path. Refer to Fig.\ref{fig_sneak_path}, if selectors are added in series to each cell, the reverse current in cell $(4,3)$ can no longer pass through due to the addition of the selector, and thereby the sneak path marked in red can be eliminated completely. However, due to the imperfections in the memory array production or maintenance, the cell selectors may fail and hence the sneak path may still occur. In this work, we follow the literature work \cite{13.sf1,14.sf2}, and assume that the selector fails \textit{i.i.d.} with probability (4,3).

Following \cite{14.sf2,17.code1}, a cell $U_{i,j}$ is affected by the sneak path if the following three conditions are met:
\begin{enumerate}
  \item $x_{i,j}=0$;
  \item There exists at least one combination of $i' \in \left\{ {1, \ldots ,m} \right\}$, $j' \in \left\{ {1, \ldots ,n} \right\}$ that induces a sneak path, defined by:
  \begin{equation*}\label{sp_condition}
    x_{i',j}=x_{i,j'}=x_{i',j'}=1;
  \end{equation*}
  \item The selector in the cell $(i',j')$ fails.
\end{enumerate}

The conditions defined above essentially assume that there are three memory cells in the sneak path parallel to the desired read path. This is because although in principle sneak paths can consist any odd number of cells,  pathes having large number of resistors have negligible contribution to the sneak path interference. Therefore, the measured resistance of a cell affected by the sneak path can be written as
\begin{equation}\label{R_sp}
  {R'_0} = {\left( {\frac{1}{{{R_0}}} + \frac{1}{{3{R_1}}}} \right)^{ - 1}}.
\end{equation}
We remark that our proposed detection scheme can be generalized to other types of sneak paths straight forwardly. In addition, following \cite{14.sf2,17.code1}, here we consider the case of a single sneak affecting a selected cell since it is the dominant case when the cell selector failure probability is small.

Therefore, the readback signal array $\mathbf{Y} = {\left[ {{y_{i,j}}} \right]_{M \times N}}$ from the ReRAM crossbar array can be written as \cite{14.sf2,17.code1}
\begin{equation}\label{readback_data}
  {y_{i,j}} = {r_{i,j}} + n_{i,j}
\end{equation}
with
\begin{equation}\label{first_step_hard}
{r_{i,j}}=
  \begin{cases}
    {{{\left( {\frac{1}{{{R_0}}} + \frac{{{e_{i,j}}}}{{{3R_1}}}} \right)}^{ - 1}}}& \text{if ${{x_{i,j}} = 0}$}\\
    {{R_1}}& \text{if ${{x_{i,j}} = 1}$}
  \end{cases},
\end{equation}
where ${e_{i,j}}$ is a Boolean random variable which takes the value of 1 if sneak path occurs at cell $U_{i,j}$, and otherwise, $e_{i,j}=0$.
Here, ${n_{i,j}}$ is an additive noise that is caused by various noise sources in the memory system such as the sensing circuit noise and cell-to-cell variations due to memory fabrication imperfection. Since the noise sources have different characteristics, we follow \cite{13.sf1,14.sf2,17.code1} and assume ${n_{i,j}}$  to be Gaussian distributed with mean $0$ and variance ${\sigma ^2}$. Our proposed detection scheme can be easily generalized to noises with other distributions, by modifying the probability density functions (PDFs) used in Section III.D. In the simulations throughout this work, we follow the literature \cite{13.sf1,14.sf2,17.code1} and set the low resistance state to be with $R_1 = 100\Omega$ and the high resistance state to be with $R_0 = 1000\Omega$. We assume the input distribution is {\it i.i.d.} Bernoulli $(1/2)$, i.e., $\Pr \left( {{x_{i,j}} = 1} \right) = \Pr \left( {{x_{i,j}} = 0} \right) =1/2$. 

\section{Belief Propagation Based Detection}
As described earlier, the occurrence of the sneak path and hence the sneak path interference is data-dependent. According to the conditions given in Section II, whether or not a high resistance cell $U_{i,j}$ will be affected by the sneak path interference depends on the resistance states of the memory cells in the same row and same column with $U_{i,j}$, and the resistance states and the selector status of the memory cells that are in the diagonal direction with respect to $U_{i,j}$. It should be noted that  the diagonal cell means that a cell is not in horizontal direction and not in vertical direction in this paper. This dependence has not been taken into consideration in most of the exiting detection schemes for ReRAM. In this section, a BP based detection scheme is proposed by exploring the dependence of the states of the memory cells that are involved in the sneak path.

\subsection{Detection of High Resistance Cells not Affected by the Sneak Path Interference}
We first notice that due to the physical characteristics of ReRAM, the high resistance value $R_0$ is much larger than the low resistance value $R_1$. Therefore, if a memory cell is corrupted by the additive noise ${n_{i,j}}$ only but not affected by the sneak path interference, during reading, the high resistance state will seldom be detected erroneously to the low resistance state.  For example, when detect the data by hard decisions without considering the sneak path interference, {\it i.e.} ${\tilde r_{i,j}} = \arg \mathop {\min }\limits_{r \in \left\{ {{R_0},{R_1}} \right\}} {\left( {{y_{i,j}} - r} \right)^2}$, with a noise variance $\sigma^2=70$, the error rate is $Q\left( {\frac{{{R_0} - \frac{{{R_1} + {R_0}}}{2}}}{\sigma }} \right) \approx 6.4404 \times {10^{ - 11}}$ which is negligible. Therefore, the cells with high resistance states can be detected firstly by
\begin{equation}\label{readback_r}
{\tilde r_{i,j}}=
  \begin{cases}
    R_0,& \text{if $\arg \mathop {\min }\limits_{r \in \left\{ {{R_0},{R_1},R{'_0}} \right\}} {\left( {{y_{i,j}} - r} \right)^2}=R_0$}\\
    R_s,& \text{otherwise}
  \end{cases},
\end{equation}
where $R_s$ denotes the uncertain resistance. It could be either a true low resistance $R_1$ stored in the memory cell, or $R'_0$ which is the resistance of a sneak path interference affected high resistance cell. The focus of our next step work is to detect whether the $R_s$ is $R_1$ or $R'_0$.

\subsection{Detection of Definite Low Resistance Cells}
As we known, one of the necessary conditions for the occurrence of sneak path interference in cell $U_{i,j}$ is that there are at least one $R_1$ in the same row and one $R_1$ in the same column of the crossbar array. Therefore, the resistance of cells with $R_s$ are definitely $R_1$ if all the cells in its same row or its same column are with high resistance states. \textcolor{black}{We denote two sets associated with the position of cell $U_{i,j}$ with $R_s$, given by}
\begin{equation}\label{R_set}
	\textcolor{black}{
{{\bf{O}}_{i,j}} = \left\{ {\left( {i,v} \right)\left| {{{\tilde r}_{i,v}} = {R_s},v \in \left[ {1,N} \right]\backslash j} \right.} \right\},}
\end{equation}
and
\begin{equation}\label{C_set}
	\textcolor{black}{
  {{\bf{C}}_{i,j}} = \left\{ {\left( {u,j} \right)\left| {{{\tilde r}_{u,j}} = {R_s},u \in \left[ {1,M} \right]\backslash i} \right.} \right\}.}
\end{equation}
\textcolor{black}{That is, the set $\mathbf{O}_{i,j}$ consists of the positions of cells with uncertain resistance $R_s$ in the same row with cell $U_{i,j}$, and the set $\mathbf{C}_{i,j}$ consists of the positions of cells with uncertain resistance $R_s$ in the same column with cell $U_{i,j}$.} If ${{\cal \mathbf{O}}_{i,j}} = \emptyset $ or ${{\cal \mathbf{C}}_{i,j}} = \emptyset $, the conditions for the cell $U_{i,j}$ to be corrupted by the sneak path interference cannot be satisfied, and hence  the resistance of cell $U_{i,j}$ definitely is $R_1$. Therefore, the decision rule of (\ref{readback_r}) is updated as
\begin{equation}\label{readback_r_up}
{\tilde r_{i,j}}=
  \begin{cases}
    R_0& \text{if $\arg \mathop {\min }\limits_{r \in \left\{ {{R_0},{R_1},R{'_0}} \right\}} {\left( {{y_{i,j}} - r} \right)^2}=R_0$}\\
    R_1& \text{if ${\mathbf{O}_{i,j}} = \emptyset$ or ${\mathbf{C}_{i,j}} = \emptyset$}\\
    R_s& \text{otherwise}
  \end{cases}.
\end{equation}

\subsection{Tanner Graph Representation of ReRAM Channel with Sneak Path Interference}
\begin{figure}
  \centering
  \includegraphics[width=3.6in, height=2.2in]{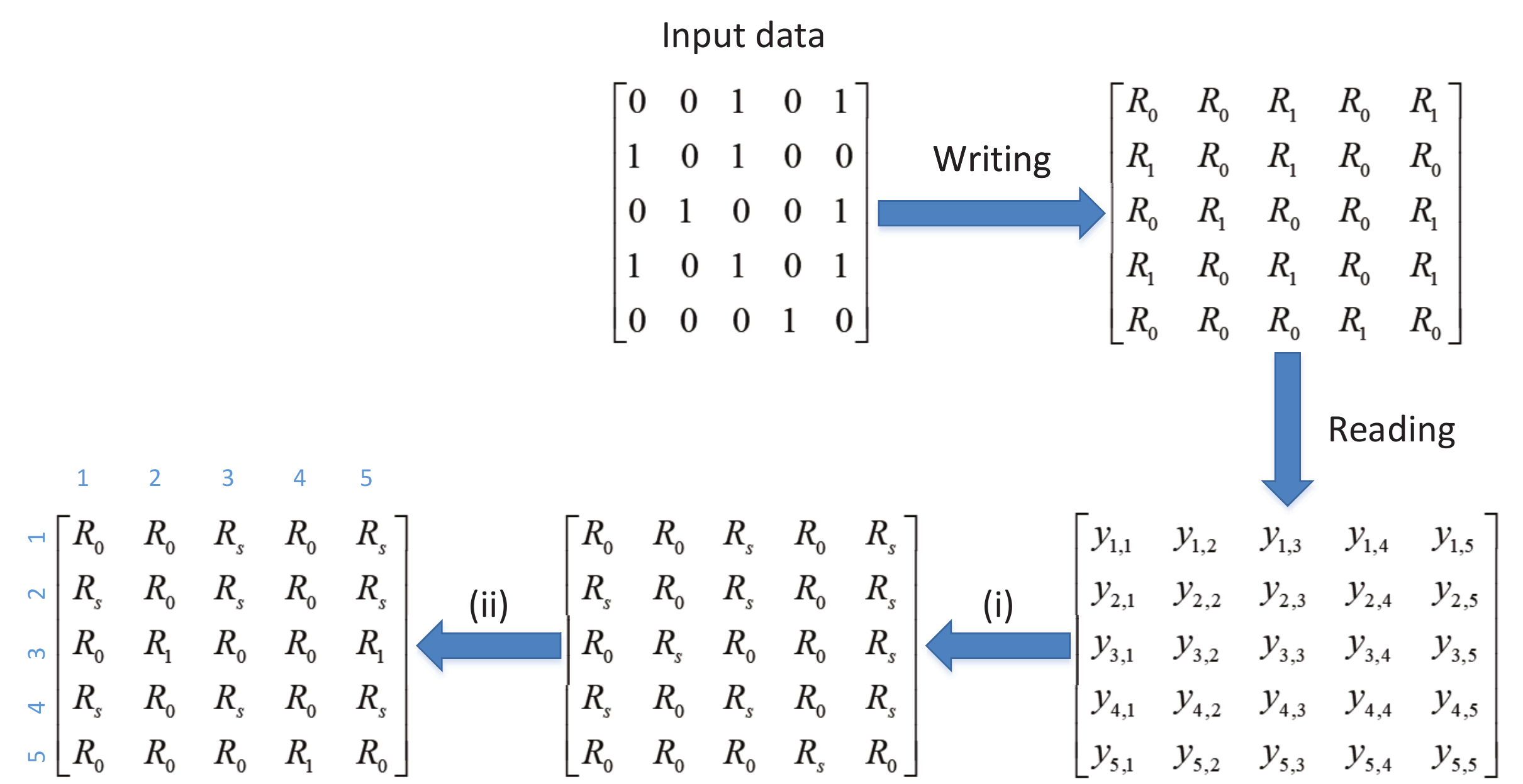}
  \caption{Example of a $5\times5$ memristor array, where the input data array is stored in the cross memory and affected by sneak path interference and noise. The readback data array is pre-detected before iteration: step (i) is to detect the high resistance $R_0$ firstly , then in step (ii), a part of low resistance $R_1$ can be detected by Eq.(\ref{readback_r_up}).}\label{fig_first_step}
\end{figure}

Except for the cells that have been determined using the decision rule given by (\ref{readback_r_up}), the remaining cells with $R_s$ are used as the nodes to construct a Tanner graph for the ReRAM channel corrupted by the sneak path interference. BP based detection can then be carried out over the constructed Tanner graph. Take a $5 \times 5$ memory array as an example which is shown in Fig. \ref{fig_first_step}.  The input user data array is stored in the resistive memory array through a writing/memory programming process. When reading the signal from the memory cell, the resistance of the memory cell may be corrupted by both the neak path interference and the additive noise. According to our proposed data detection given by (\ref{readback_r_up}), Step (i) is to detect the high resistances $R_0$ that are not affected by the sneak path interference first based on the Euclidean distance calculation. After that in step (ii), memory cells that are definitely with the low resistance $R_1$ are detected. The remaining cells with the uncertain resistance $R_s$ will be used to construct a Tanner graph. % so that they can be detected through the proposed BP detection.

%Since memory cells with the uncertain resistance $R_s$ could be a cell corrupted by the sneak path interference, and it also could be a cell with a faulty selector. We now consider the latter case where a cell $U_{i,j}$ with $R_s$ whose cell selector may be faulty.

\begin{figure}
  \centering
  \includegraphics[width=2.8in, height=1.8in]{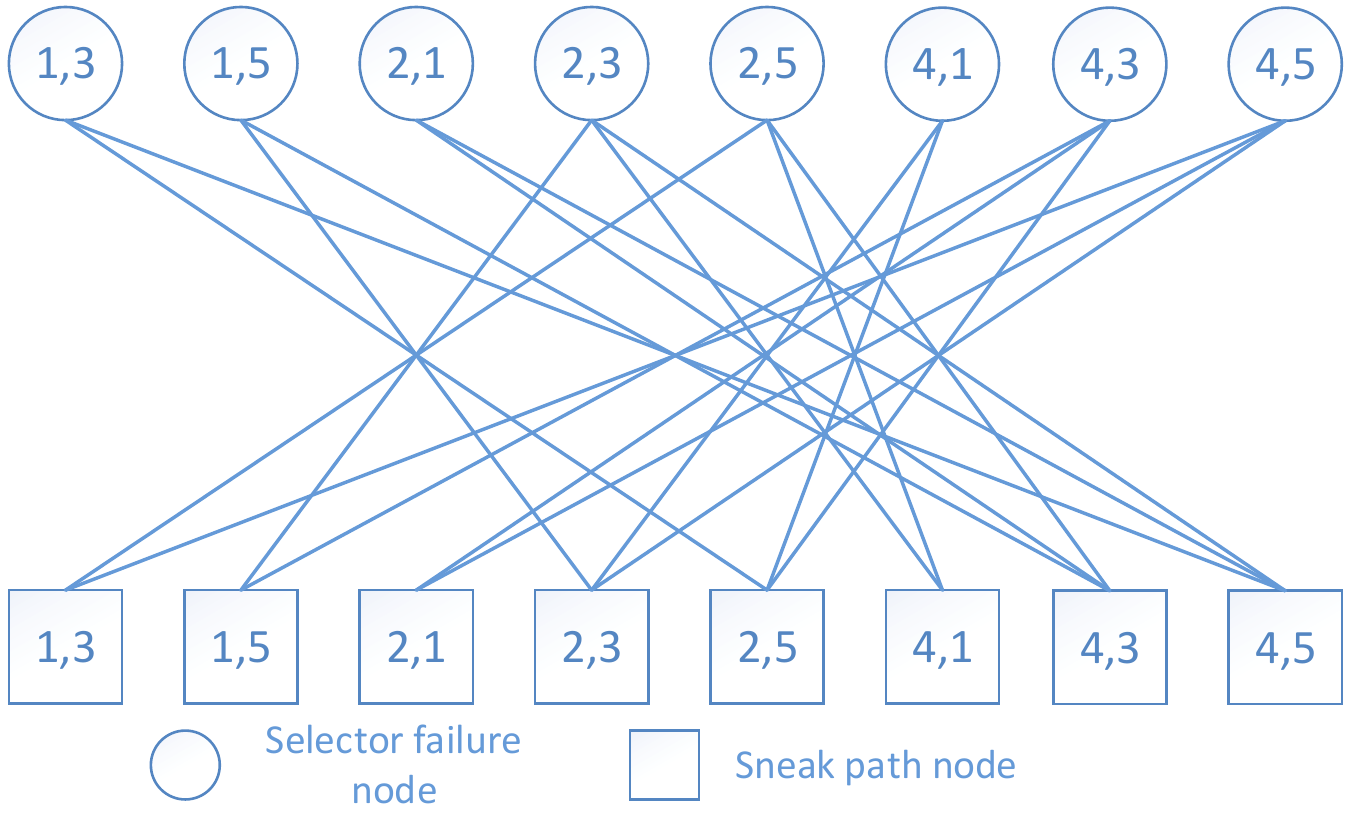}
  \caption{Tanner graph constructed based on the last data array in Fig.\ref{fig_first_step}, where each cell $U_{i,j}$ with resistance $R_s$ is set as the nodes in the graph. The circle represents SFN and the square represents SPN. An edge joins a SFN to a SPN if the SFN has a probability to cause the sneak path interference to SPN.}\label{fig_secend_step}
\end{figure}

The details for constructing the Tanner graph are as follows. First, for a cell $U_{i,j}$ with resistance $R_s$, we define another set ${\mathbf{D}_{i,j}}$ as
\begin{equation}\label{D_set}
	\textcolor{black}{
   {{\bf{D}}_{i,j}} = \left\{ {\left( {u,v} \right)\left| {{{\tilde r}_{u,v}} = {R_s},\left( {u,j} \right) \in {{\bf{C}}_{i,j}},\left( {i,v} \right) \in {{\bf{O}}_{i,j}}} \right.} \right\},}
\end{equation}
which denotes the positions of diagonal cells with respect to $U_{i,j}$ with resistance $R_s$, with $1 \le u \le M, 1 \le v \le N,u \ne i,v \ne j$. Since the resistance of cell $U_{i,j}$ is $R_s$ which could be $R_1$, the selector in $U_{i,j}$ may fail, and the resistance of cells in ${\mathbf{O}_{i,j}}$ and ${\mathbf{C}_{i,j}}$ may be $R_1$, the cells in ${\mathbf{D}_{i,j}}$ may be affected by the sneak path interference that caused by the faulty selector in $U_{i,j}$. Therefore, the cell $U_{i,j}$ is denoted as the selector failure node (SFN) and the cells in ${\mathbf{D}_{i,j}}$ are denoted as sneak path nodes (SPNs) for SFN $(i,j)$. \textcolor{black}{An edge joins a SFN to a SPN if the SFN will cause the sneak path of SPN when the conditions of sneak path are satisfied.} Due to the uncertainty of $R_s$, each cell with resistance $R_s$ is a SFN, and at the same time, it could also be a SPN. Hence both the numbers of SFNs and SPNs are equal to the number of cells with resistance $R_s$. \textcolor{black}{In the last matrix of Fig.\ref{fig_first_step}, there are eight cells with $R_s$, and these cells are set to be eight SFNs and eight SPNs in the Tanner graph illustrated by Fig.\ref{fig_secend_step}. We take the cell $U_{4,5}$ as an example, whose diagonal sets ${\mathbf{D}_{(4,5)}} = \left\{ {\left( {1,3} \right),\left( {2,1} \right),\left( {2,3} \right)} \right\}$ . We regard $U_{4,5}$ as a SFN, therefore, the cells $U_{1,3}$, $U_{2,1}$ and $U_{2,3}$ are the SPNs for $U_{4,5}$.}
\subsection{Message Passing for Belief Propagation based Detection}
After the Tanner graph for the ReRAM channel is constructed, the next step is to carry out the BP based detection.
We denote the event that the selector in cell $U_{i,j}$ fails by $SF_{i,j}$, and the event that the cell $U_{m,n}$ is affected by the sneak path interference by $SP_{m,n}$. The message passed from the SFN $U_{i,j}$ to the connected SPNs is the probability of $SF_{i,j}$ with the readback signal array being $\mathbf{Y}$. The message passed from the SPN $U_{m,n}$ to the connected SFNs is the probability of $SP_{m,n}$. The output of the SPN $U_{m,n}$, ({\it i.e.}, the output of the BP detection) is the probability of the event that $r_{m,n}=R_1$ when the readback data is $y_{m,n}$.

\begin{figure}
  \centering
  \includegraphics[width=1.8in, height=1.6in]{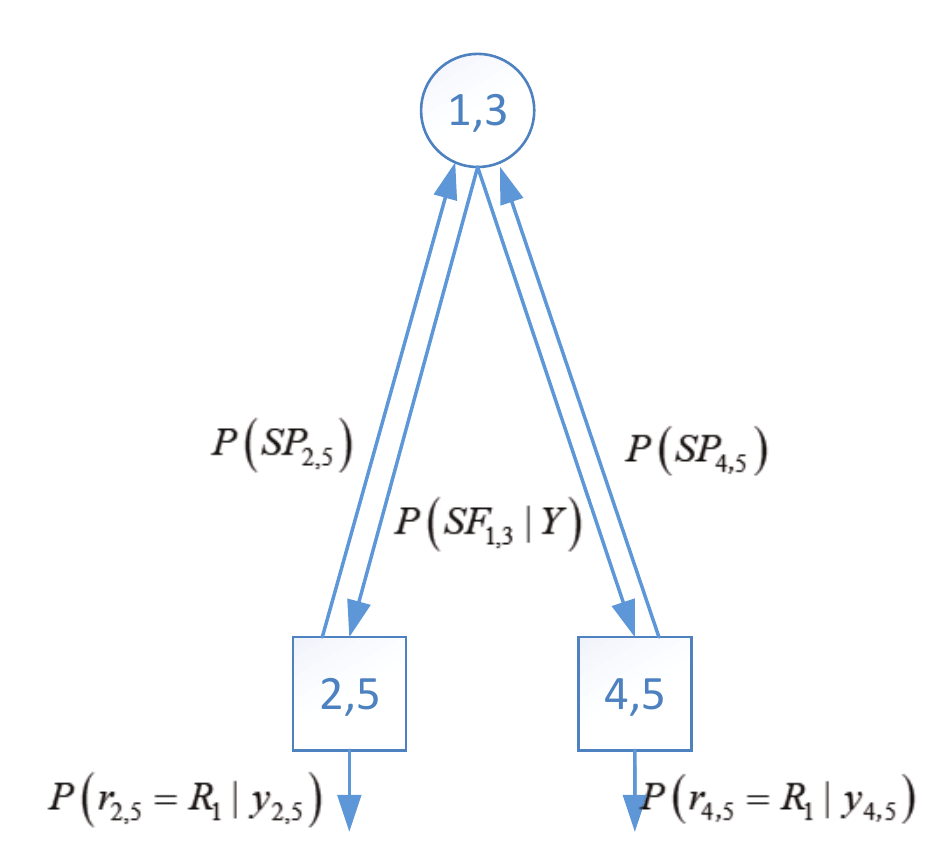}
  \caption{The message propagation among a SFN $(1,3)$ and its SPNs $(2,5)$ and $(4,5)$. The message from SFN $(1,3)$ to the connected SPNs is the probability of $SF_{1,3}$ when the readback signal array is $\mathbf{Y}$. The messages from SPNs $(2,5)$ and $(4,5)$ to SFN $(1,3)$ are the probabilities of $SP_{2,5}$ and $SP_{4,5}$, respectively, and the outputs of the SPNs are the probabilities of the events that $r_{2,5}=R_1$ and $r_{4,5}=R_1$ when the readback signals are $y_{2,5}$ and $y_{4,5}$, respectively.}\label{fig_onenode}
\end{figure}

The corresponding message propagation is shown by Fig. \ref{fig_onenode}.
In particular, the message from the SFN $U_{i,j}$ to the SPNs $U_{m,n}$ is given by
\begin{equation}\label{SFN}
\begin{split}
  P\left( {S{F_{i,j}}|\mathbf{Y}} \right) & = \frac{{P\left( {\mathbf{Y}|S{F_{i,j}}} \right)P\left( {S{F_{i,j}}} \right)}}{{P\left( \mathbf{Y} \right)}}\\
    & \mathop  = \limits^{(a)} \frac{{P\left( {{\mathbf{Y}_{{D_{i,j}}}}|S{F_{i,j}}} \right)P\left( {{\mathbf{Y}_{D_{i,j}^\mathbf{C}}}|S{F_{i,j}}} \right)}}{{P\left( {{\mathbf{Y}_{{D_{i,j}}}}} \right)P\left( {{\mathbf{Y}_{D_{i,j}^\mathbf{C}}}} \right)}}P\left( {S{F_{i,j}}} \right)\\
    & \mathop  = \limits^{(b)} \frac{{P\left( {{\mathbf{Y}_{{D_{i,j}}}}|S{F_{i,j}}} \right)}}{{P\left( {{\mathbf{Y}_{{D_{i,j}}}}} \right)}}P\left( {S{F_{i,j}}} \right)\\
    & \textcolor{black}{\mathop  = \limits^{(c)} \frac{{\prod\limits_{\left( {u,v} \right) \in {D_{i,j}}\backslash \left( {m,n} \right)} {P\left( {{y_{u,v}}|S{F_{i,j}},{{\tilde r}_{u,v}} = {R_s}} \right)} }}{{\prod\limits_{\left( {u,v} \right) \in {D_{i,j}}\backslash \left( {u,v} \right)} {P\left( {{y_{u,v}}|{{\tilde r}_{u,v}} = {R_s}} \right)} }}P\left( {S{F_{i,j}}} \right)},
\end{split}
\end{equation}
where ${\mathbf{Y}_{{D_{i,j}}}}$ denotes the readback signals of cells in $D_{i,j}$ and ${\mathbf{Y}_{D_{i,j}^\mathbf{C}}}$ denotes the readback signals of cells not in $D_{i,j}$. It is reasonable to assume that ${\mathbf{Y}_{{D_{i,j}}}}$ and ${\mathbf{Y}_{D_{i,j}^\mathbf{C}}}$ are independent, and hence the equality (a) is satisfied. The equality in (b) holds since $SF_{i,j}$ only affects the cells in $D_{i,j}$. That is, $P\left( {{\mathbf{Y}_{D_{i,j}^\mathbf{C}}}|S{F_{i,j}}} \right) = P\left( {{\mathbf{Y}_{D_{i,j}^\mathbf{C}}}} \right)$. The equality (c) follows from the assumption that the cells in $D_{i,j}$ are independent with each other. In (\ref{SFN}), the term $P\left( {{y_{u,v}}|{{\tilde r}_{u,v}} = {R_s}} \right)$ can be obtained as

\begin{equation}\label{y_mn}
  \begin{split}
    P&\left( {{y_{u,v}}|{{\tilde r}_{u,v}} = {R_s}} \right) \\
     = &P\left( {{y_{u,v}}|{r_{u,v}} = {R_1}} \right)P\left( {{r_{u,v}} = {R_1}|{{\tilde r}_{u,v}} = {R_s}} \right) \\
        & + P\left( {{y_{u,v}}|{r_{u,v}} = {{R'}_0}} \right)P\left( {{r_{u,v}} = {{R'}_0}|{{\tilde r}_{u,v}} = {R_s}} \right) \\
      = & \varepsilon(S{P_{u,v}})\phi \left( {{y_{u,v}},{R_1}} \right)  + [1-\varepsilon(S{P_{u,v}})]\phi \left( {{y_{u,v}},{R'_0}} \right),
  \end{split}
\end{equation}
where
\begin{equation}\label{Gaussion}
  \phi \left( {{y},{R_x}} \right) = \frac{1}{{\sqrt {2\pi } \sigma }}{e^{ - \frac{{{{\left( {y - {R_x}} \right)}^2}}}{{2{\sigma ^2}}}}},
\end{equation}
and $\varepsilon (S{P_{m,n}})$ is the probability of the event that the resistance of cell $U_{m,n}$ is $R'_0$ if ${\tilde r_{m,n}} = R_s$
\begin{equation}\label{num_sp_node}
  \varepsilon (S{P_{u,v}}) = \frac{{\left( {1 - q} \right)P\left( {S{P_{u,v}}} \right)}}{{\left( {1 - q} \right)P\left( {S{P_{u,v}}} \right) + q}}.
\end{equation}
Similarly,
\begin{equation}\label{y_mn_sf}
  \begin{split}
    P\left( {{y_{u,v}}|S{F_{i,j}},{{\tilde r}_{u,v}} = {R_s}} \right)  = &\varepsilon (S{P_{u,v}}|S{F_{i,j}})\phi \left( {{y_{u,v}},{R_1}} \right) \\
        &  + [1 - \varepsilon (S{P_{u,v}}|S{F_{i,j}})]\phi \left( {{y_{u,v}},{R'_0}} \right)
  \end{split}
\end{equation}
By defining a function
\begin{equation}\label{fun}
\begin{split}
 f\left( {m,n,u,v} \right) = & P\left( {{r_{m,v}} = {R_1}|{y_{m,v}}} \right)\\
 &P\left( {{r_{u,n}} = {R_1}|{y_{u,n}}} \right)P\left( {{r_{u,v}} = {R_1}|{y_{u,v}}} \right),
 \end{split}
\end{equation}
the message from the SPN $U_{m,n}$ to SFNs $U_{i,j}$ can be expressed as
\begin{equation}\label{SPN}
  \textcolor{black}{P\left( {S{P_{m,n}}} \right) = 1 - \prod\limits_{\left( {u,v} \right) \in {D_{m,n}}\backslash \left( {i,j} \right)} {\left[ {1 - f\left( {m,n,u,v} \right)P\left( {S{F_{u,v}}|{\bf{Y}}} \right)} \right]}}, 
\end{equation}
and
\begin{equation}\label{SPN_sf}
	\textcolor{black}{
  \begin{split}
     P&\left( {S{P_{m,n}}|S{F_{i,j}}} \right) =   1 - f\left( {m,n,i,j} \right) \\
      & \times \prod\limits_{\left( {u,v} \right) \in {D_{m,n}}\backslash \left( {i,j} \right)} [{1 - } f\left( {m,n,u,v} \right)P\left( {S{F_{u,v}}|\mathbf{Y}} \right)].
  \end{split}}
\end{equation}
Apart from propagating messages to SFNs, SPNs also need to output the probability that the resistance of cell $U_{m,n}$ is $R_1$, given by
\begin{equation}\label{one}
 \begin{split}
   & P\left( {{r_{m,n}} = {R_1}|{y_{m,n}}} \right) \\
   & = \frac{{P\left( {{y_{m,n}}|{r_{m,n}} = {R_1}} \right)P\left( {{r_{m,n}} = {R_1}} \right)}}{\sum\limits_{{r_{u,v}} = \left\{ {{R'_0},{R_1}} \right\}} {P\left( {{y_{m,n}}|{r_{u,v}}} \right)} P\left( {{r_{u,v}}} \right)} \\
     &  = \frac{{\phi \left( {{y_{m,n}},{R_1}} \right)\left[ {1 - \varepsilon \left( {S{P_{m,n}}} \right)} \right]}}{{\phi \left( {{y_{m,n}},{R_1}} \right)\left[ {1 - \varepsilon \left( {S{P_{m,n}}} \right)} \right] + \phi \left( {{y_{m,n}},{{R'}_0}} \right)\varepsilon \left( {S{P_{m,n}}} \right)}}
 \end{split}
\end{equation}
The BP detection iteratively computes an approximation of: the probability of selector failure in each cell; the  probability that the cell is affected by sneak path interference; and the probability that the resistance of the cell is $R_1$. Before the iteration of messages, the probability that each SPN is affected by the sneak path interference is initialized as
\begin{equation}\label{SP_init}
\begin{split}
  P&\left( {S{P_{init}}} \right) = E\left[ {\frac{{\sum\limits_{i = 1}^M {\sum\limits_{j = 1}^N {{e_{i,j}}} } }}{{MN\left( {1 - q} \right)}}} \right] \\
    & = P\left( {{e_{i,j}}|{x_{i,j}} = 0} \right)\\
    & = 1 - \sum\limits_{u = 0}^{M - 1} {\sum\limits_{v = 0}^{N - 1} {C_{M - 1}^uC_{N - 1}^v} } {q^{u + v}}{\left( {1 - q} \right)^{M - 1 - u + N - 1 - v}}\\
    &\qquad \qquad \qquad \quad \times {\left( {1 - {p_{sf}}q} \right)^{uv}},
\end{split}
\end{equation}
where $p_{sf}$ is the a priori probability of the selector failure, and $P(SF_{i,j}) = p_{sf}$ for each cell in the first iteration. After a certain number of iterations, the SPNs output the estimation of the resistance for each cell to be $R_s$, given by

\begin{equation}\label{estimation}
{r_{m,n}}=
  \begin{cases}
    R_1 & \text{if ${P\left( {{r_{mn}} = {R_1}|{y_{m,n}}} \right) \ge 0.5}$}\\
    {R'_0}& \text{else}.
  \end{cases}
\end{equation}
By further combining with (\ref{readback_r_up}), the estimation of data array is given by
\begin{equation}\label{est_data}
{\tilde x_{i,j}}=
  \begin{cases}
    0 & \text{if $\tilde r_{i,j}=R_0$ or $\tilde r_{i,j}=R'_0$}\\
    1 & \text{if $\tilde r_{i,j}=R_1$}.
  \end{cases}
\end{equation}
The above proposed BP detection algorithm is summarized in Algorithm.1.
\begin{figure}[t]
  \centering
  \includegraphics[width=2.4in, height=2.0in]{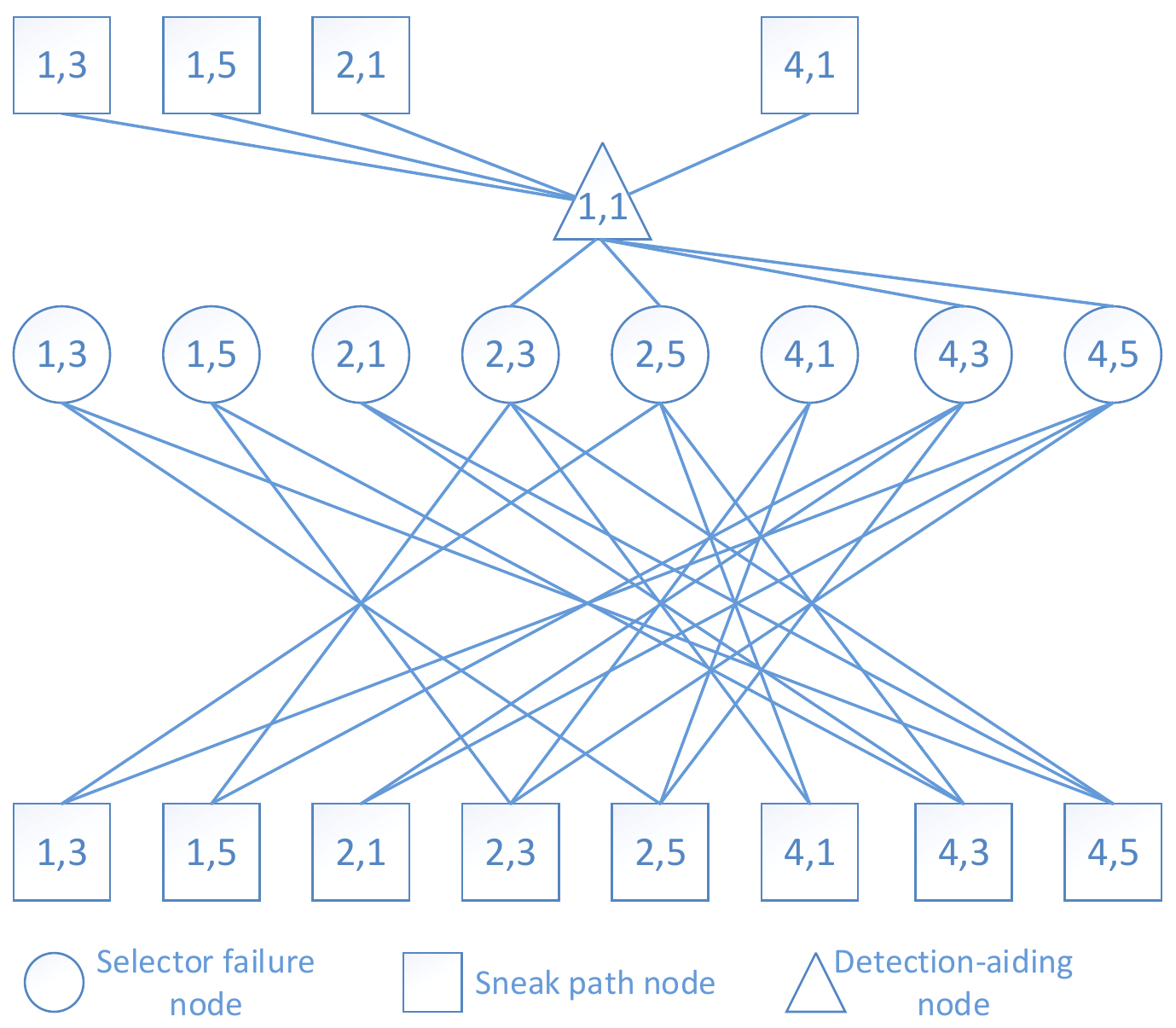}
  \caption{Tanner graph for improved BP detection, where the DANs are connected with SFNs and the messages propagated inside the graph become more accurate.}\label{fig_ZN}
\end{figure}

\renewcommand{\algorithmicrequire}{\textbf{Input:}}%Input
\renewcommand{\algorithmicensure}{\textbf{Output:}}%Output
\begin{algorithm}
	\begin{spacing}{1.4}
\caption{BP detection for ReRAM system}
\begin{algorithmic}[1]
\Require the readback signal \textbf{Y}, the variance of noise $\sigma ^2$
\Ensure the estimation of data $\tilde X$

\For {$(i,j)=(1,1)$ to $(M,N)$}

$// Decision$
    \State Calculate ${\tilde r_{i,j}}$ by Eq.(\ref{readback_r})

\EndFor
\For {$U_{i,j}=U_{1,1}$ to $(U_{M,N}$}
$// Select\;  nodes$
    \If {${\tilde r_{i,j}} = {R_s}$}
        \State Set ${\mathbf{O}_{i,j}}$, ${\mathbf{C}_{i,j}}$ by Eq.(\ref{R_set}) and Eq.(\ref{C_set})
        \If {${\mathbf{O}_{i,j}} = \emptyset $} or {${{\mathbf{C}}_{i,j}} = \emptyset $}
            \State ${\tilde r_{i,j}} = {R_1}$
        \Else
            \State Push ${\tilde r_{i,j}}$ to SPN
        \EndIf
    \EndIf
\EndFor
\For {each {$U_{i,j}$} in $SPN$}
$// Construct\;  Tanner \; graph$
    \State Set ${\mathbf{D}_{i,j}}$ by Eq.(\ref{D_set}) 
\EndFor
\State Set $I_{max}$, $SFN = SPN$
\State Initialize $SP$ by Eq.(\ref{SP_init}) and $p_{sf}$
\For {$i=1$ to $I_{max}$}
$// Iteration$
    \For {each {$U_{i,j}$} in $SFN$}
        \State Update $P(SF_{i,j})$ by Eq.(\ref{SFN}) \textcolor{black}{(that is calculate by Eq.(\ref{y_mn})-(\ref{y_mn_sf}))}
    \EndFor
    \For {each {$U_{m,n}$} in $SPN$}
        \State Update $P(SP_{m,n})$ by Eq.(\ref{SPN})
        \State Calculate $P\left( {{r_{m,n}} = {R_1}|{y_{m,n}}} \right)$ by Eq.(\ref{one}) \textcolor{black}{(that is calculate by Eq.(\ref{Gaussion})-(\ref{num_sp_node}))}
    \EndFor
\EndFor
\State Estimate the data array $\tilde X$ by Eq.(\ref{est_data})
\State \Return{The estimation of data: $\tilde X$}
    \end{algorithmic}
\end{spacing}
\end{algorithm}

\subsection{Improved Belief Propagation based Detection}
We notice that if the the resistance of the cell $U_{u,v}$, the diagonal cell with respect to $U_{i,j}$, is $R_0$, the greater the probability of $\tilde r_{u,j}=R_1$ and $\tilde r_{i,v}=R_1$, the less likely the selector fails at cell $U_{i,j}$. Therefore the cells with $R_0$ can provide side information to improve the performance of the BP detection. We define the detection-aiding set for cell $U_{i,j}$ as
\begin{equation}\label{Z_set}
	\textcolor{black}{
{{\bf{Z}}_{i,j}} = \left\{ {\left( {u,v} \right)\left| {{{\tilde r}_{u,v}} = {R_0},\left( {u,j} \right) \in {{\bf{O}}_{i,j}},\left( {i,v} \right) \in {{\bf{C}}_{i,j}}} \right.} \right\},
}
\end{equation}
and the cells in detection-aiding set is regarded as the DANs for cell $U_{i,j}$. The message pass from the DANs to SFN is given by
\begin{equation}\label{SFN_zero}
  \begin{split}
     &P\left( {S{F_{i,j}}|{{\tilde r}_{{Z_{i,j}}}} = {R_0}} \right)  \\
      & = \frac{{P\left( {{{\tilde r}_{{Z_{i,j}}}} = {R_0}|S{F_{i,j}}} \right)P\left( {S{F_{i,j}}} \right)}}{{P\left( {{{\tilde r}_{{Z_{i,j}}}} = {R_0}} \right)}} \\
       & = P\left( {S{F_{i,j}}} \right)\prod\limits_{\left( {u,v} \right) \in {Z_{i,j}}} 1 - f\left( {u,v,i,j} \right)
  \end{split}
\end{equation}
The function $f(\cdot)$ in (\ref{SFN_zero}) can be initialized by (\ref{fun}) and the initialization of $P\left( {S{F_{i,j}}} \right)$ is $p_{sf}$. The probabilities of the corresponding cells being with low resistance states are propagated form the SPNs to DANs for calculating the function $f(\cdot)$.

With the aid of DANs, the messages propagated between the nodes of Tanner graph become more accurate. Correspondingly, Fig. \ref{fig_secend_step} is updated to Fig. \ref{fig_ZN}, and (\ref{SFN}) is updated as
\begin{equation}\label{1}
P\left( {S{F_{i,j}}|\mathbf{Y}} \right) = P\left( {S{F_{i,j}}} \right)\cdot \alpha \cdot \prod\limits_{\left( {u,v} \right) \in {Z_{i,j}}} 1 - f\left( {u,v,i,j} \right) ,
\end{equation}
where
\begin{equation*}\label{alpha}
  \alpha = \frac{{\prod\limits_{\left( {m,n} \right) \in {D_{i,j}}} {P\left( {{y_{m,n}}|S{F_{i,j}},{{\tilde r}_{m,n}} = {R_s}} \right)} }}{{\prod\limits_{\left( {m,n} \right) \in {D_{i,j}}} {P\left( {{y_{m,n}}|{{\tilde r}_{m,n}} = {R_s}} \right)} }}.
\end{equation*}

\begin{figure}
  \centering
  \includegraphics[width=3.4in, height=2.8in]{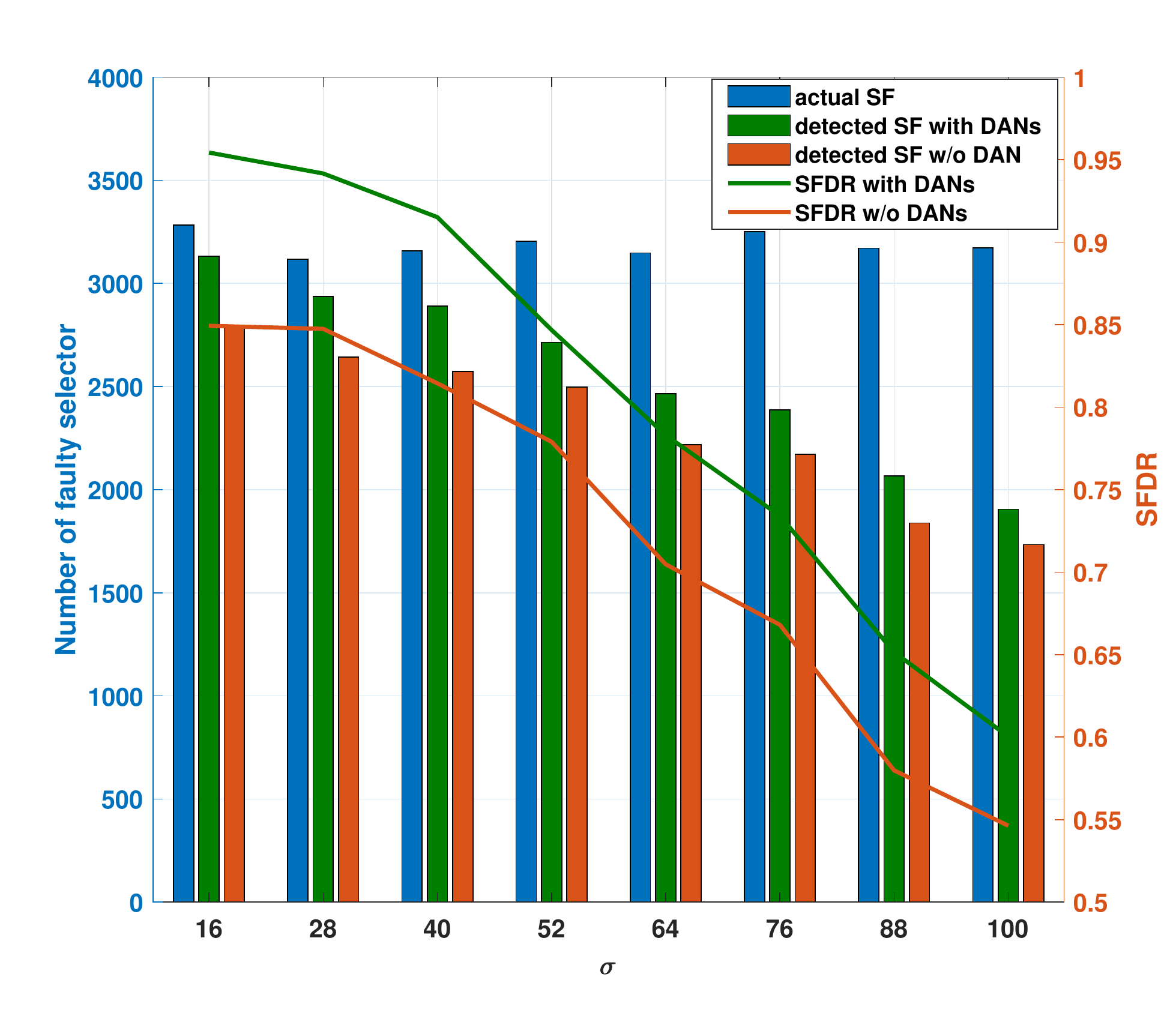}
  \caption{The comparison between the number of actual faulty selectors and the number of detected faulty selectors, where the memory size is $M\times N=8\times8$, $p_{sf}=0.001$.}\label{fig_001_8r}
\end{figure}

\begin{figure}
  \centering
  \includegraphics[width=3.4in, height=2.8in]{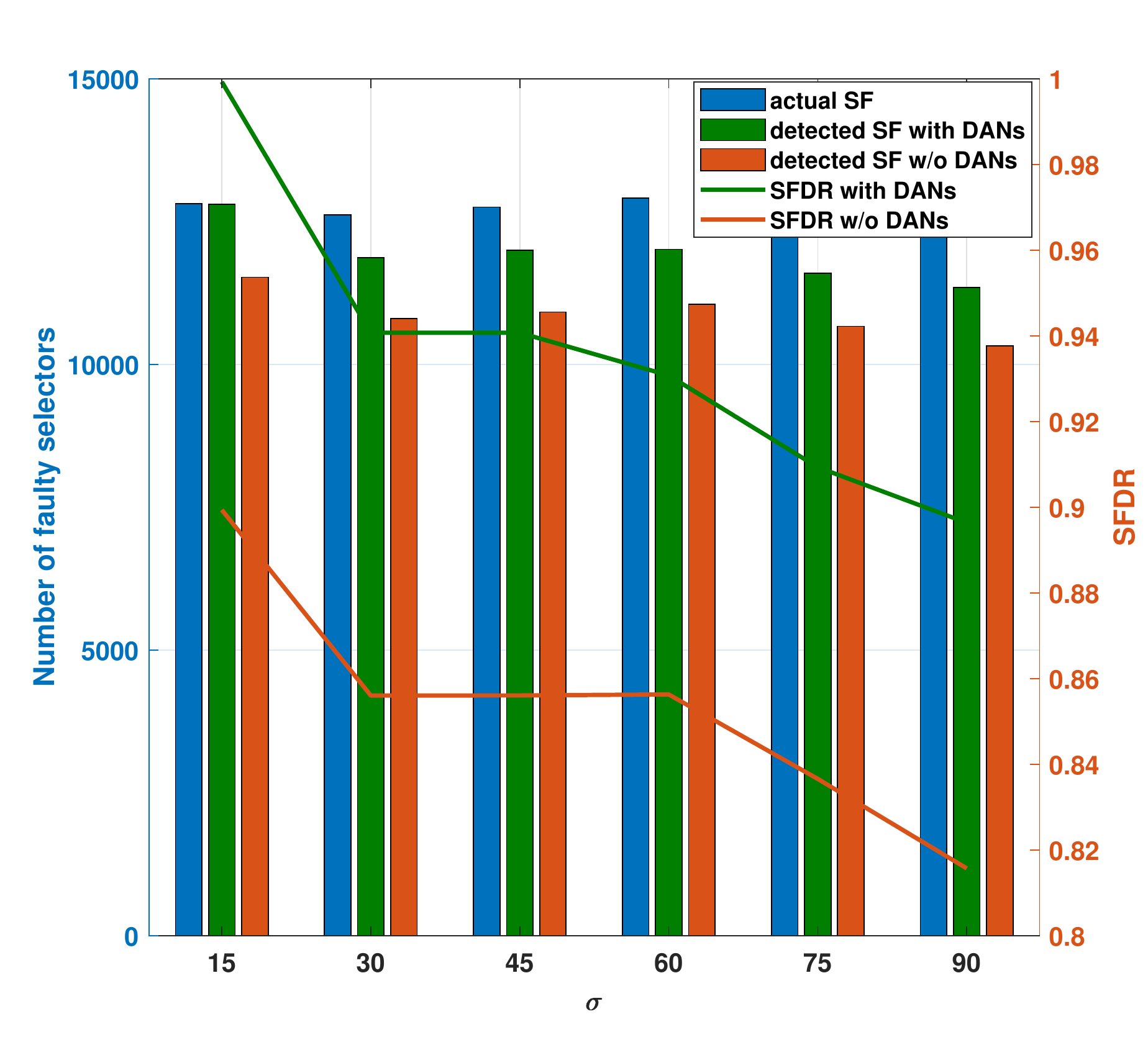}
  \caption{The comparison between the number of actual faulty selectors and the number of detected faulty selectors, where the memory size is $M\times N=16\times16$, $p_{sf}=0.001$.}\label{fig_001_16r}
\end{figure}

Next, to verify the above proposed BP detector, we simulate the selector failure detection rate (SFDR) which is defined as
\begin{equation}\label{sfdr}
  SFDR = \frac{{{n_{\det }}}}{{{n_{sf}}}},
\end{equation}
where $n_{det}$ is the number of faulty selectors with $P(SF_{i,j})>0.99$ after the proposed BP detection, and $n_{sf}$ is the actual number of failed selectors. Since the failure of selector is the key factor that causes the sneak path interference, the accuracy of faulty selector detection is critical for combatting the sneak path interference. The simulation results are illustrated in Fig. \ref{fig_001_8r} and Fig. \ref{fig_001_16r}, for array sizes of $M\times N=8\times 8$ and $16\times 16$, respectively. It can be observed that the proposed BP detector can effectively detect the faulty selectors, and the improved BP detector further improves the SFDR. Furthermore, the system with a larger array size can detect the faulty selector more accurately since in this case more nodes can be incorporated in the detection iterations.

\section{Belief Propagation based Joint Detection and Decoding with Polar Codes}

\subsection{Belief Propagation based Joint Detection and Decoding}
The BP detector proposed in Section III can detect the cells affected by sneak path interference effectively, and generate the corresponding soft information. Apart from the sneak path interference, the ReRAM channel is also corrupted by the channel noise. In this work, we further cascade the
graph of the BP detector constructed in the previous section with that of the BP decoder and present a BP based joint detection and decoding scheme to further improve the reliability of ReRAM.
%\subsection{Coding and Decoding Model}

\begin{figure}
  \centering
  \includegraphics[width=3.0in, height=1.6in]{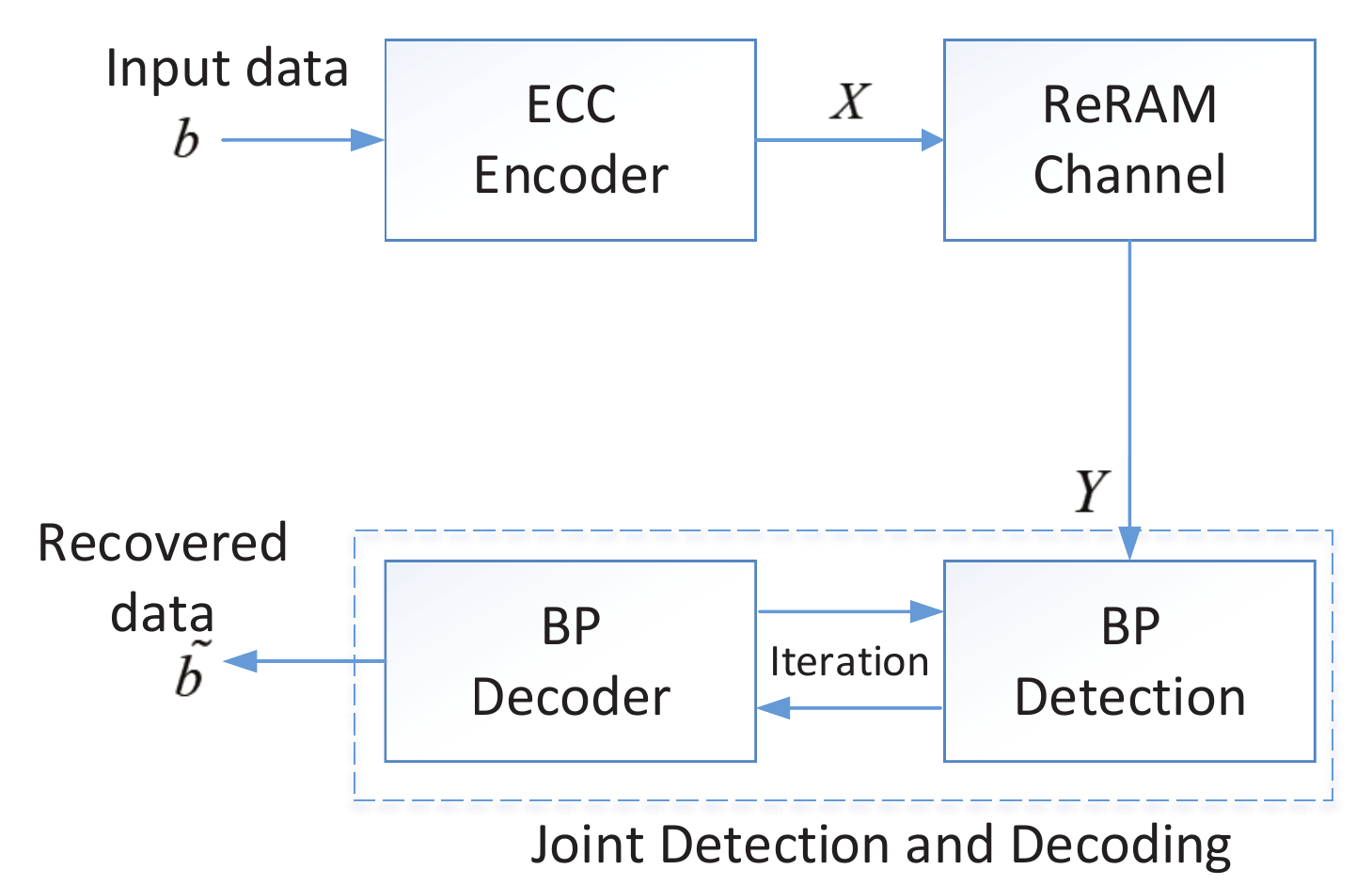}
  \caption{BP based joint detection and decoding system for ReRAM.}\label{fig_model}
\end{figure}

A system model for the proposed joint detection and decoding scheme is illustrated in Fig. \ref{fig_model}. The input user data, denoted by $\textbf{\textit{b}} = \left\{ {{b_1},{b_2},...,{b_K}} \right\}$, is encoded by an ECC encoder into the codeword $\textbf{X} \in {\left\{ {0,1} \right\}^{M \times N}}$ before being stored into the resistive crossbar memory. The code rate is $R_{ECC}=K/(MN)$. Correspondingly, the ECC decoder, chosen to be a BP decoder, is cascaded with the BP detector to achieve a joint detection and decoding. Messages propagate between the BP detector and decoder iteratively.

%\subsection{Joint Detection}
Although both the LDPC codes and polar codes can be adopted in the proposed BP based joint detection and decoding, we consider the polar codes in this work, which are the first provably capacity-achieving codes with low encoding and decoding complexity \cite{20.Polar}. \textcolor{black}{What's more, the code lengths in this work are relatively short, and polar codes have better error-rate performance when code length is short as demonstrated in the 5G wireless standard.} Compared with the serial SC decoder, the BP decoder for polar codes requires short decoding delay and can achieve high throughput due to its inherent high parallelism, which is more suitable for high-speed memories such as the ReRAM.

\begin{figure}
  \centering
  \includegraphics[width=3.5in, height=2.3in]{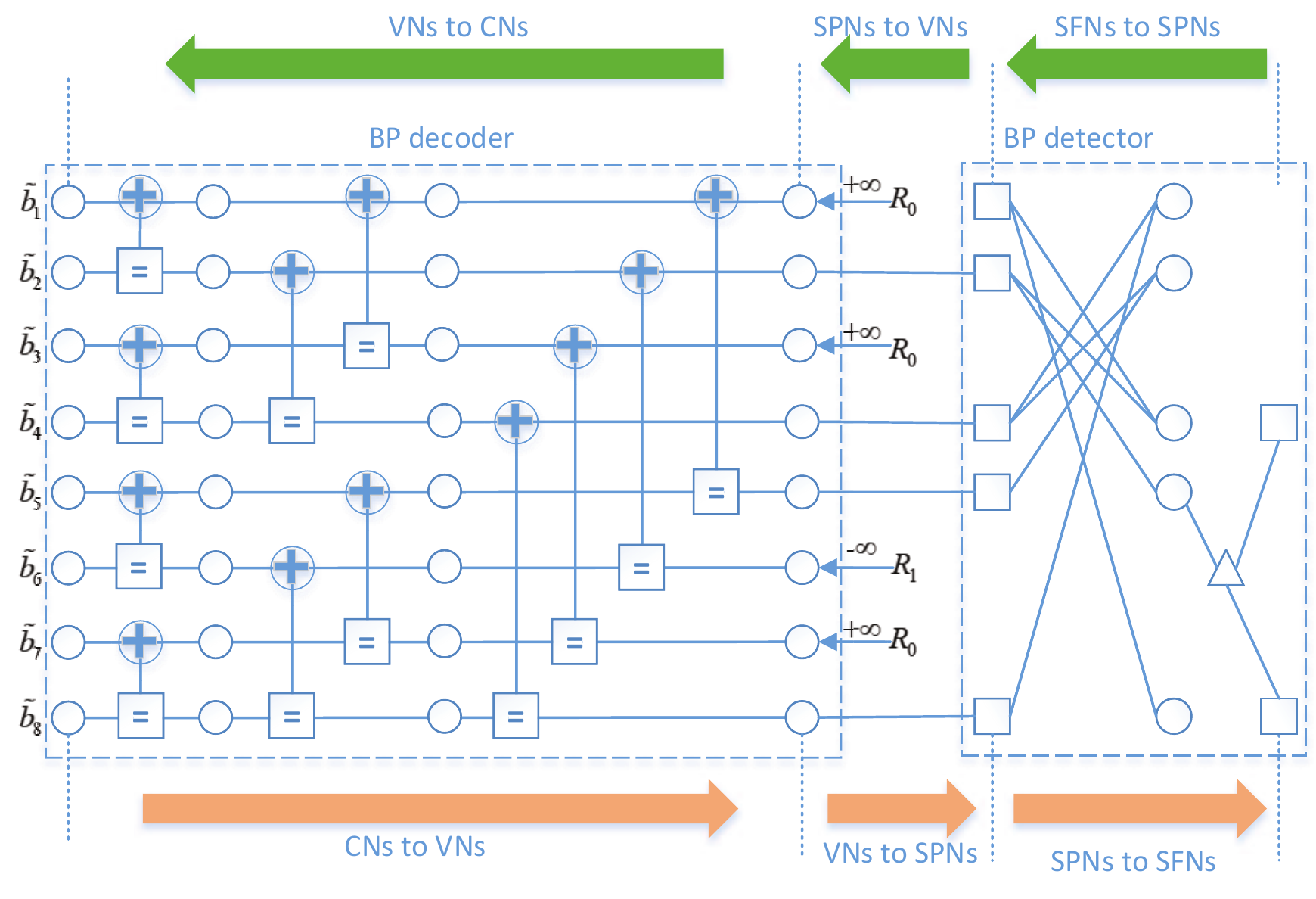}
  \caption{Example of a joint detector and the polar decoder, where the SPNs of BP detection are connected with the VN of polar decoder. In each iteration, the SPNs output the extrinsic LLR to the VNs. The extrinsic LLR is used as a priori LLR to VNs processing.}\label{fig_iteration}
\end{figure}

Since in the Tanner graph of the BP decoder, reliability information is passed iteratively along the edges that connect the check nodes (CNs) and variable nodes (VNs), the VNs can naturally be connected with the SPNs of the BP detector described in the previous section to form a joint graph. As shown in
Fig. \ref{fig_iteration}, the output of the SPN of the BP detector is the probability that the resistance of cell is $R_1$, which can be converted into LLR and send to the input of the VN, given by
\begin{equation}\label{SPN2VN}
  LL{R_{i,j}} = \log \frac{{1 - P\left( {{r_{i,j}} = {R_1}|{y_{i,j}}} \right)}}{{P\left( {{r_{i,j}} = {R_1}|{y_{i,j}}} \right)}}.
\end{equation}
The SPN $U_{i,j}$ is connected with the corresponding variable node $VN_t$ of the BP decoder, with $t = (i-1)*N+j$. Within each joint detection/decoding iteration, the SPNs output the extrinsic LLRs to the VNs, which are used as a priori LLRs for the VN processing. For VNs that are not matched with the SPNs of the BP detector, the LLRs are fixed as
\begin{equation}\label{fixed_LLR}
LL{R_{\left( {i - 1} \right)*j + j}} = \left\{ {\begin{array}{*{20}{l}}
\infty &{{\rm{if }} \quad {{\tilde r}_{i,j}} = {R_0}}\\
{ - \infty }&{{\rm{if }} \quad {{\tilde r}_{i,j}} = {R_1}}
\end{array}} \right.,
\end{equation}
and LLRs of these VNs are not updated during according iteration. After half a round of iteration from the BP detector to the BP decoder, the soft information of the CN outputs is passed to the VNs and then to the SPNs of the BP detector to complete the second half of the iteration. The BP detection is used to detect the faulty selector and the sneak path interference, and the BP decoding is used to correct the errors caused by channel additive noise, and hence the joint detection and decoding can detect the sneak path interference and correct the errors caused by channel noise simultaneously in each iteration. After a certain number of iterations, the VNs of the BP decoder in the joint detector and decoder output the reliable LLR for each ECC coded bit.
The estimated user data bit from the joint detector and decoder is given by
\begin{equation}\label{joint_data}
  {\tilde b_{i,j}} = \left\{ {\begin{array}{*{20}{l}}
0 &{{\rm{if }} \quad LL{R_{^{i,j}}} \geq 0}\\
1 &{{\rm{if }} \quad LL{R_{^{i,j}}} < 0}
\end{array}} \right.,
\end{equation}

\subsection{Polar Codes Construction for the ReRAM channel}
 A key step of polar construction is to determine the information-bit indexes set and frozen-bit indexes set from the polarized sub-channels. Existing construction methods such as the DE algorithm and PW algorithm are optimized for the SC decoding of polar codes over the symmetric channels. However, the ReRAM channel is asymmetric due to the data-dependent sneak path interference. Hence the decoding performance with the conventional polar construction will be limited. Therefore, in this sub-section, polar code construction tailored for the BP based joint detector and decoder over the ReRAM channel is proposed based on the GenA.

\begin{figure}
  \centering
  \includegraphics[height=3.5in]{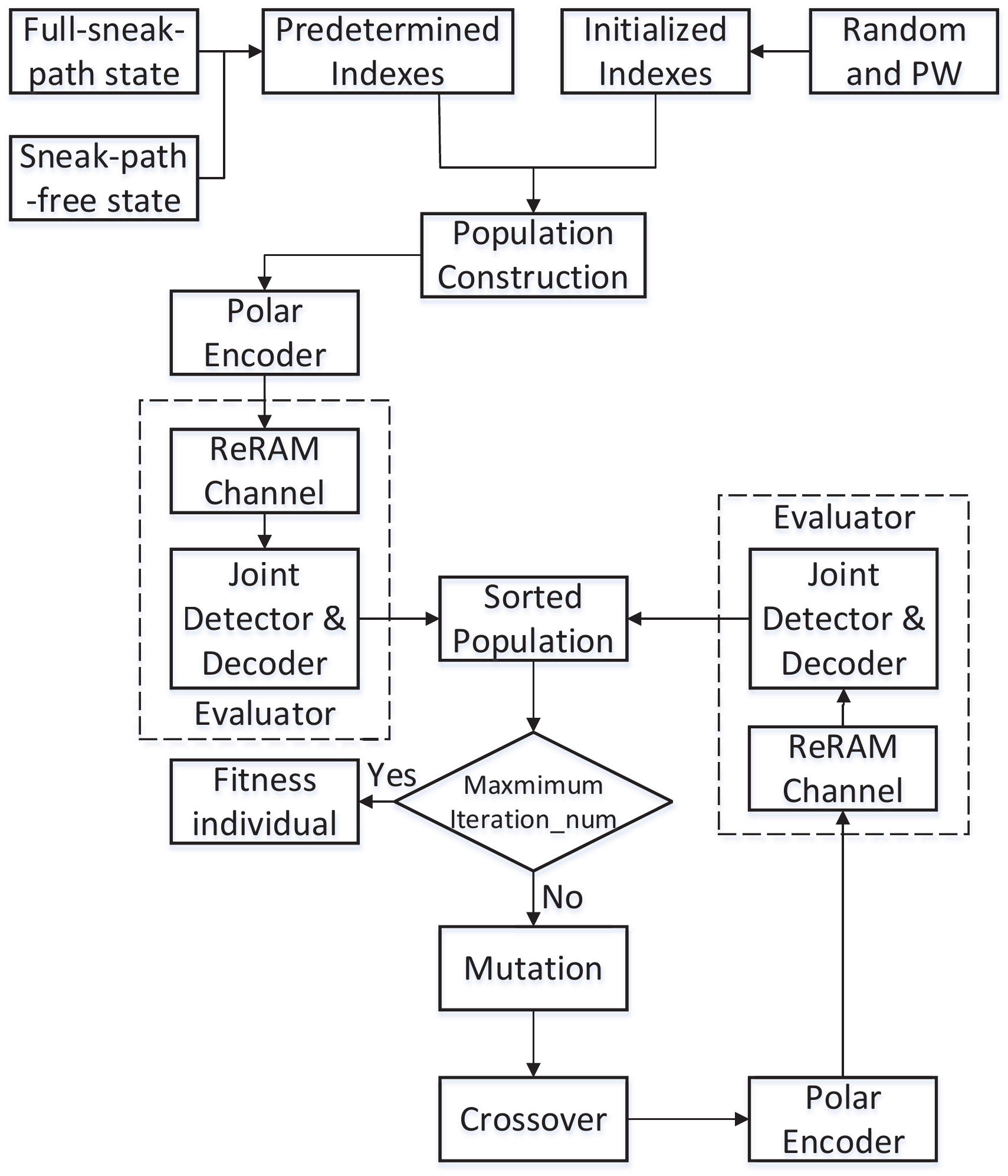}
  \caption{Polar codes construction based on GenA for joint detection and decoding over the ReRAM channel.}\label{polar_construction}
\end{figure}

On the other hand, based on the characteristics of the ReRAM channel, we can define two extreme channel states: the full-sneak-path state where we can find some sub-channels that are always reliable, and the sneak-path-free state where we can find some sub-channels that are always unreliable. Therefore, we can predetermine some information-bit indexes before applying the GenA to find the other information-bit indexes. This will reduce the overall complexity for the code construction, and may also make the GenA converge faster. The entire frame work of the polar code construction is depicted in Fig. \ref{polar_construction}.

\subsubsection{Predetermined Information-bit Indexes}
Some information-bit indexes can be predetermined by considering the following two extreme states of the ReRAM channel.

(a)\ {Full-sneak-path state}:
For the full-sneak-path (FSP) state, we assume that all the cells with high resistances are affected by the sneak path interference. Hence, the ReRAM channel is turned to a symmetric AWGN channel with a low signal-to-noise ratio (SNR), where polar codes can be constructed by DE or the PW algorithm for this channel. The channel quality of the FSP is worse than that of the actual ReRAM channel, and hence the sub-channels that are reliable enough in the FSP state must be reliable for the actual ReRAM channel. In this state, the sub-channel indexes that are reliable are denoted as $\mathcal{A}_{pre}$. Let $\left[ N \right] = \left\{ {1,2,...,N} \right\}$ and $\beta < 1/2$ be a fixed positive constant. The $\mathcal{A}_{pre}$ is obtained by
\begin{equation*}\label{A_pre}
  {\mathcal{A}_{pre}} \buildrel \Delta \over = \left\{ {i \in \left[ N \right]:Z\left( {{W_i}} \right) < {2^{ - {N^{\beta _1}}}}/N} \right\},
\end{equation*}
where ${Z\left( {{W_i}} \right)}$ is the Bhattacharyya parameter of sub-channel $W_i$ that can be obtained using the DE or PW algorithm.

(b)\ {Sneak-path-free state}:
On the contrary, for the sneak-path-free (SPF) state, we assume that all cells with high resistances are not affected by the sneak path interference. Hence, the ReRAM channel is turned to a symmetric AWGN channel with a high SNR. The channel quality of the SPF is better than that of the actual ReRAM channel, and thereby the sub-channels that are unreliable in the FSP state must be unreliable for the actual ReRAM channel. In this state, the sub-channel indexes that are unreliable are denoted as $\mathcal{F}_{pre}$, which can be obtained by
\begin{equation*}\label{F_pre}
  {\mathcal{F}_{pre}} \buildrel \Delta \over = \left\{ {i \in \left[ N \right]:Z\left( {{W_i}} \right) > {2^{ - {N^{{\beta _2}}}}}/N} \right\}.
\end{equation*}

Because the FSP and SPF are two extreme channel states of ReRAM, the sub-channel indexes $\mathcal{A}_{pre}$ are always reliable and the sub-channel indexes $\mathcal{F}_{pre}$ are always unreliable irrespective of how the sneak path interference affects each individual cell. The reliability of the sub-channel set $\mathcal{A}_{pre}^\mathcal{C} \cap \mathcal{F}_{pre}^\mathcal{C}$, denoted by $\mathcal{Q}$, however, is dependent on the impact of sneak path interference on different memory cells. In the following, we determine the information-bit indexes in set $\mathcal{Q}$ by using the GenA.

\begin{figure}
  \centering
  \includegraphics[width=2.5in]{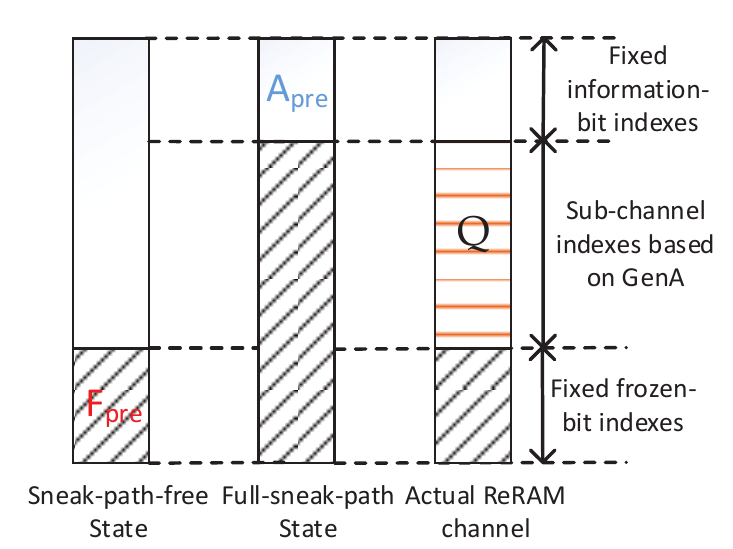}\\
  \caption{The sub-channel indexes of different states, where the sets $\mathcal{A}_{pre}$ and $\mathcal{F}_{pre}$ are the predetermined indexes sets.}\label{sub_channel_indexes}
\end{figure}

\subsubsection{Information-bit Indexes Determined by GenA}

The GenA mimics the evolution in the nature where populations of offsprings from their parents keep evolving and only the fittest offsprings can survive \cite{21.GenA}.
The task of polar code construction can be viewed as an optimization problem, searching for the optimum information set that has the minimum (possible) cost
function. This optimization problem can be solved using the GenA.

(a)\ {Population}:
In GenA, the population is a collection of candidate sets that may be sampled at any arbitrary search instance. In this work, the population $P = \left\{ {{\mathcal{A}_i}} \right\}$ consists of the different information-bit index sets from $\mathcal{Q}$, each set being denoted as $\mathcal{A}_i$, $i = 1,2,...,S$. Here, $S$ is the total number of sets in the population, and the length of $\mathcal{A}_i$ is $N*R_{ECC}-|\mathcal{A}_{pre}|$, with $|\mathcal{A}_{pre}|$ being the size of $\mathcal{A}_{pre}$. The initialized population includes two parts, the first part are the sets that are generated by the conventional PW algorithm so as to converge faster for the GenA, and the other part consists of sets that are selected randomly so as to provide a larger search space. The individual set in the population will be ranked by the evaluator.

(b)\ {Evaluator}:
In GenA, the evaluator is a fitness function which provides each individual in the population a guideline of the direction for achieving the optimal solution.
In this work, we select the block error rate (BLER) as the cost function for the optimization conducted over the  ReRAM channel. Therefore, the ReRAM channel and our proposed BP based joint detector and decoder are combined to form the evaluator for the polar codes construction. Each codeword constructed by the individuals in the population and the offsprings produced from parents are passed through this evaluator, and the BLER that can be obtained based on sampling techniques, such as the Monte-Carlo (MC) method, is used to evaluate the fitness and rank the individuals in the population.

(c)\ {Update offspring and iteration}:
The information-bit index sets in the population are ranked by the evaluator according to ascending order of BLERs, and two of them $\mathcal{A}_i$ and $\mathcal{A}_j$ will then be selected to be the parents. The parents are selected according to a probability distribution ${e^{ - \delta i}}/\sum\limits_{j = 1}^M {{e^{ - \delta j}}} $, where $\delta$ is an adjustment factor. This probability distribution ensures that the candidates with smaller BLERs have a larger chance to be selected to be the parents. We take three steps to generate the new population. First, the mutation in GenA is to provide more diversity for the population. Some random information-bit indexes in $\mathcal{A}_i$ (and $\mathcal{A}_j$) are replaced by the bit indexes in $\mathcal{A}_i^\mathcal{C}$ ($\mathcal{A}_j^\mathcal{C}$). The polar codes constructions after mutation are written as $\mathcal{A}_{im}$ and $\mathcal{A}_{jm}$. Then the crossover, as the most important step for GenA, guarantees that the GenA evolutes towards the optimal solution. For the ReRAM-tailored construction, the crossover is to merge the information-bit indexes from the parents $\mathcal{A}_{im}$ and $\mathcal{A}_{jm}$ to the offspring construction, denoted by $\mathcal{A}_o$.
Finally, $\mathcal{A}_o$ is evaluated by the evaluator and inserted into an appropriate location in the population according to the evaluated BLERs. After being updated, the last one in the population will be discarded.

The whole GenA iterates the above steps until the maximum iteration number is reached. Then the first construction $\mathcal{A}_{fir}$ (as the fittest construction) in the population is combined with $\mathcal{A}_{pre}$ to form the ultimate polar codes construction $\mathcal{A} = {\mathcal{A}_{pre}} \cup {\mathcal{A}_{fir}}$. Compared with conventional polar construction methods, the above described GenA based construction method not only provides better BLER performance for the ReRAM channel, it also has lower complexity for code construction.

\section{Simulation Results}

Computer simulations are carried out to validate the above proposed BP based joint detection and decoding system for ReRAM.
\textcolor{black}{In particular, we first evaluate the error rate performance of the proposed BP detector, and compare it with that of the threshold scheme in \cite{14.sf2} and the elementary signal estimator (ESE) detector presented in the latest work of \cite{17.code1}.} In \cite{17.code1}, the authors define the fraction of the sneak path affected cells over the total number of high resistance cells in the array as the sneak path rate $\epsilon$, and the ESE can generate an estimation of $\epsilon$ over a memory array as $\tilde \epsilon = \frac{{n_{R'_0}}}{{n_{R'_0} + n_{{R_0}}}}$, where $n_x$ is the total number of $y_{i,j}$ whose hard decision is $x$. Next, the ESE detector calculates the LLR for each bit in the data array as
\begin{equation}\label{ESE_next}
  \begin{split}
  L\left( {{x_{i,j}}|{y_{i,j}},\epsilon} \right) & = \log \frac{{\Pr \left( {{y_{i,j}}|{x_{i,j}} = 0,\epsilon} \right)}}{{\Pr \left( {{y_{i,j}}|{x_{i,j}} = 1,\epsilon} \right)}}\\
  &  = \log \frac{{\epsilon \phi \left( {{y_{i,j}},R{'_0}} \right) + \left( {1 - \epsilon} \right)\phi \left( {{y_{i,j}},{R_0}} \right)}}{{\phi \left( {{y_{i,j}},{R_1}} \right)}},
  \end{split}
\end{equation}
based on which the hard decision for $x_{i,j}$ can be made. On the other hand, to get further insight on how the knowledge of failed selector can affect the data detection performance, we also include a lower bound of the error rate performance in the performance comparison. It is obtained by simulations by an ideal assumption that the BP detector knows exactly which cell's selector is failed.

%The ReRAM system without coding is simulated firstly to verify the effective of BP detector. Two ReRAM systems with the memory array size of $M\times N=8\times 8$ and $16\times 16$ are considered in this paper. For each size memory array,  And the number of iteration for BP detector is set as $I_{max}=15$.
\begin{figure} [t]
	\centering
	\includegraphics[width=3.2in, height=2.8in]{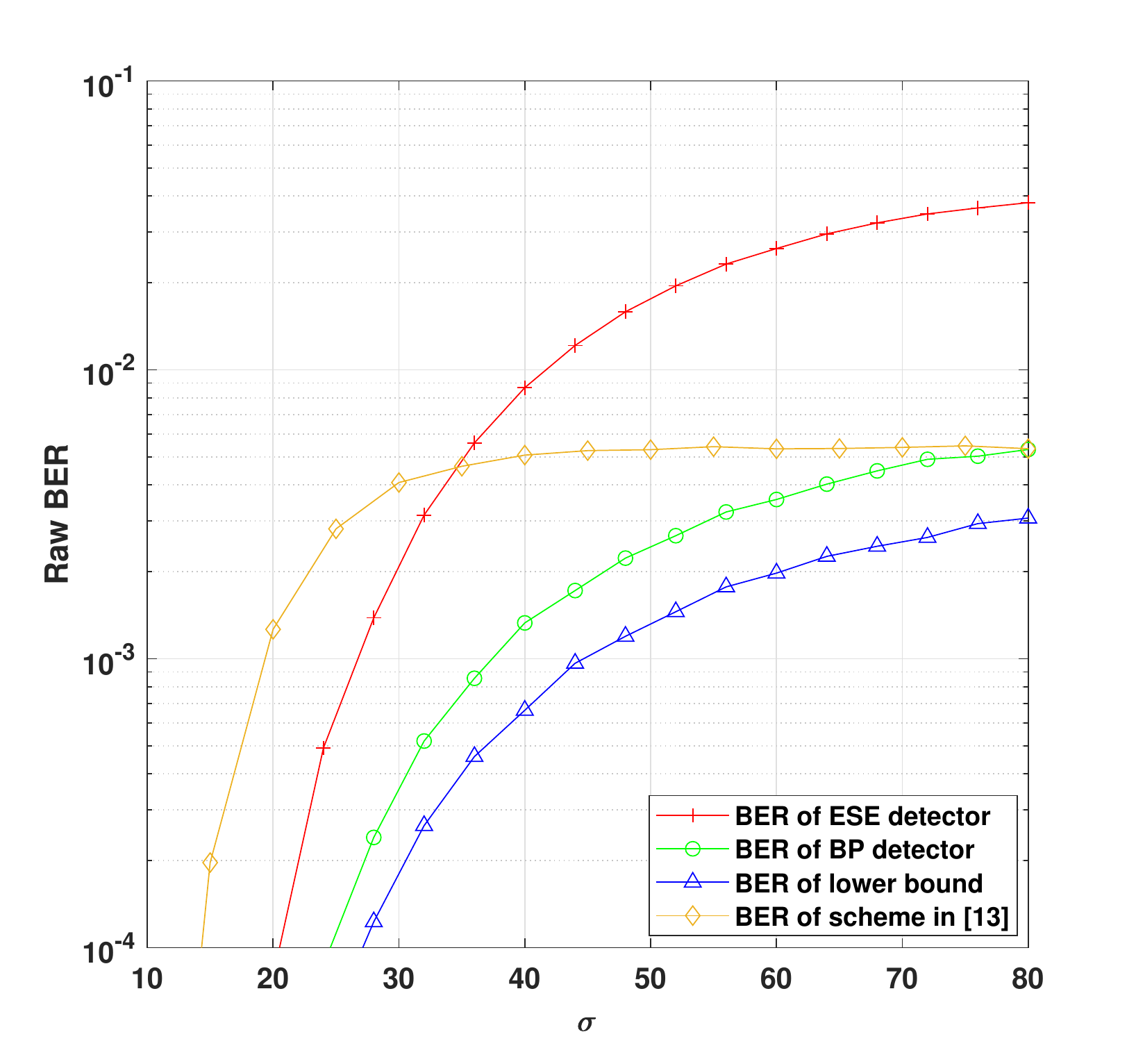}
	\caption{\textcolor{black}{Raw BER comparison of the BP detector, ESE detector \cite{17.code1}, threshold scheme \cite{14.sf2}, and the lower bound, with $M\times N=8\times8$ and $p_{sf}=0.001$.}}\label{fig_001_8}
\end{figure}
\begin{figure}[t]
	\centering
	\includegraphics[width=3.2in, height=2.8in]{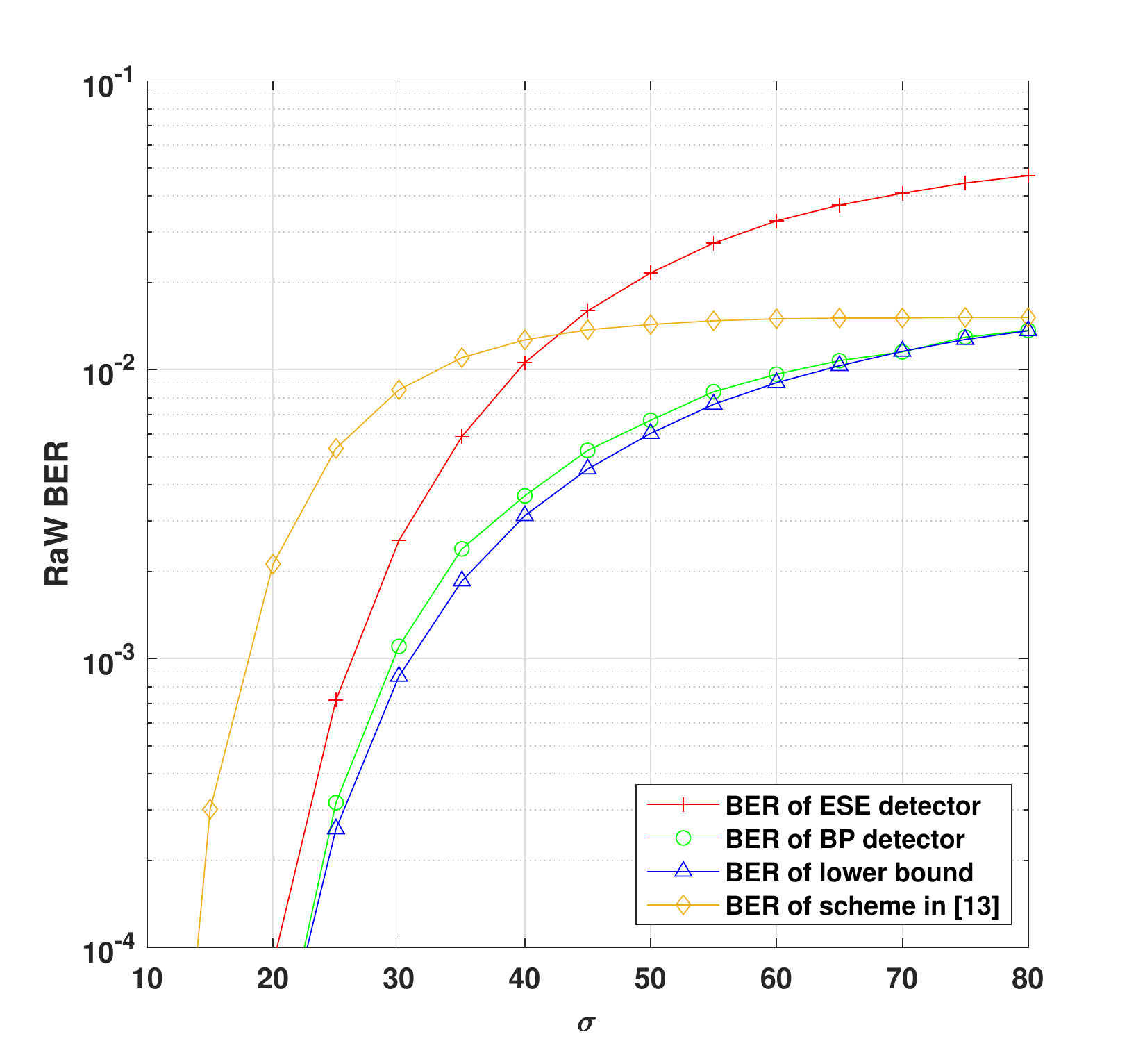}
	\caption{\textcolor{black}{Raw BER comparison of the BP detector, ESE detector \cite{17.code1}, threshold scheme \cite{14.sf2}, and the lower bound, with $M\times N=16\times16$ and $p_{sf}=0.001$.}}\label{fig_001_16}
\end{figure}

\begin{figure}[t]
	\centering
	\includegraphics[width=3.2in, height=2.8in]{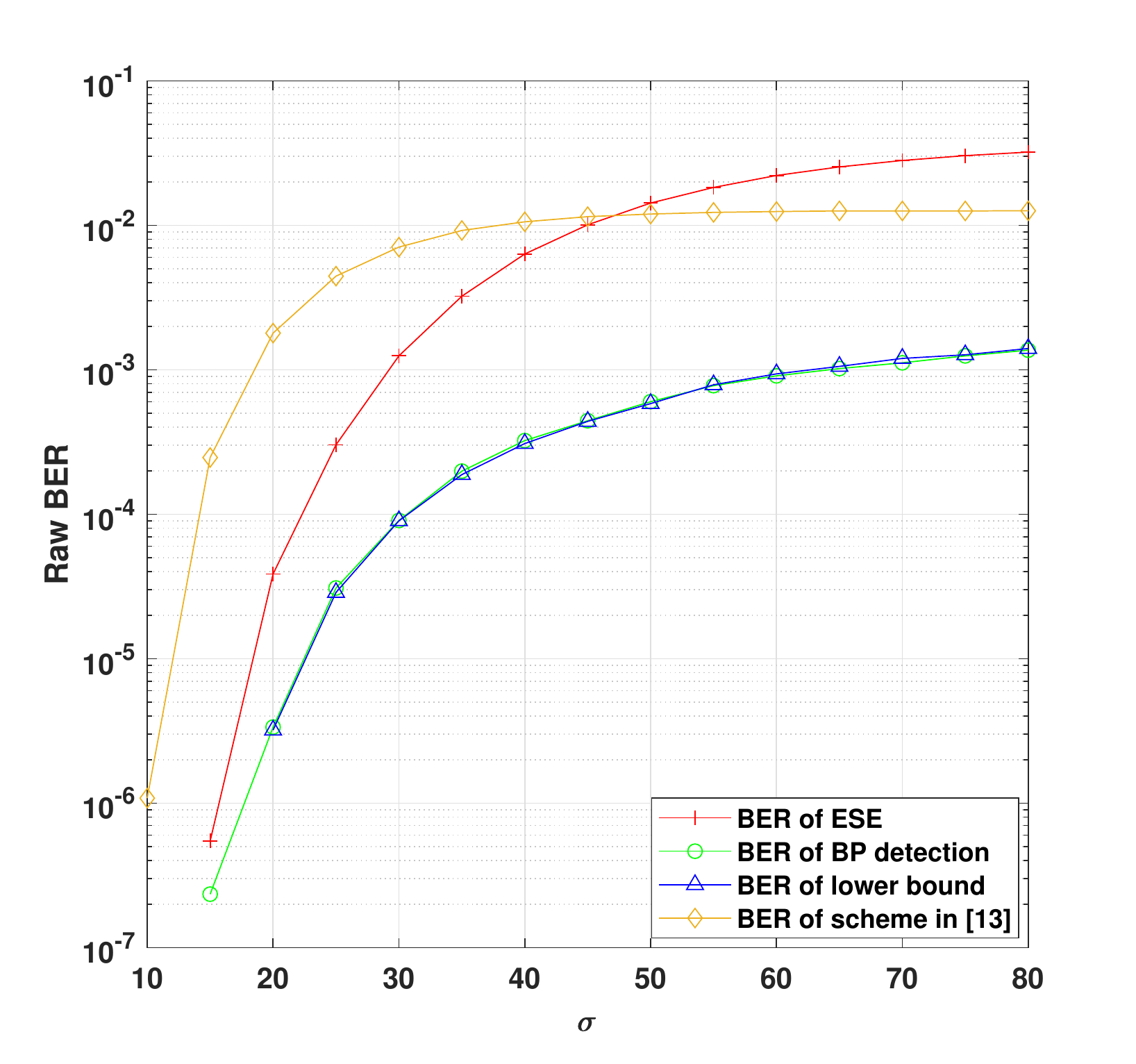}
	\caption{\textcolor{black}{Raw BER comparison of the BP detector, ESE detector \cite{17.code1}, threshold scheme \cite{14.sf2}, and the lower bound, with $M\times N=16\times16$ and $p_{sf}=0.0001$.}}\label{fig_0001_16}
\end{figure}

The BER and BLER comparisons are shown by Fig. \ref{fig_001_8} and Fig. \ref{fig_0001_16}, for memory array sizes of $M\times N=8\times 8$ and $16\times 16$, respectively. The number of iteration for the BP detector is set as $I_{max}=15$. \textcolor{black}{Observe that the proposed BP detector performs significantly better than the threshold scheme and ESE detector in terms of both the BER and BLER}, and its performance is close to the corresponding lower bound. The performance gain is attributed to the iterative nature of the BP detector which can achieve better detection based on the dependence between the faulty selectors and the cells affected by the sneak path interference. These results are consistent with the simulation results illustrated by Fig. \ref{fig_001_8r} and Fig. \ref{fig_001_16r}, which shows that the BP detector can effectively detect the faulty selectors. It is further observed that with a larger array size of $M\times N=16\times 16$, the performance of the BP detector almost overlaps with the lower bound. This is because with a larger array, there are more nodes available to provide more information about the selector failures and the sneak path interference.

In addition, from the figures, we observe that the performance gain of the BP detector is greater at medium to high noise power regions than that at the low noise power region.  The reason is that the larger amount of noise at the medium to high noise power region leads to more confusion between the cells that are truly with low resistance states with those that are affected by the sneak path interference, for the ESE detector which ignores the data dependent property and regards the sneak path interference as random noise. Since the working region for the channel detectors of data storage systems is $\rm BER=10^{-2}$ to $\rm BER=10^{-3}$, we can conclude that the BP detector can significantly improve the raw BER of the system. This will bring substantial performance gain after ECC decoding, which is demonstrated by our subsequent simulations shown below.

%\subsection{Simulation of ReRAM System With Coding}
\begin{figure}
  \centering
  \includegraphics[width=3.2in, height=2.8in]{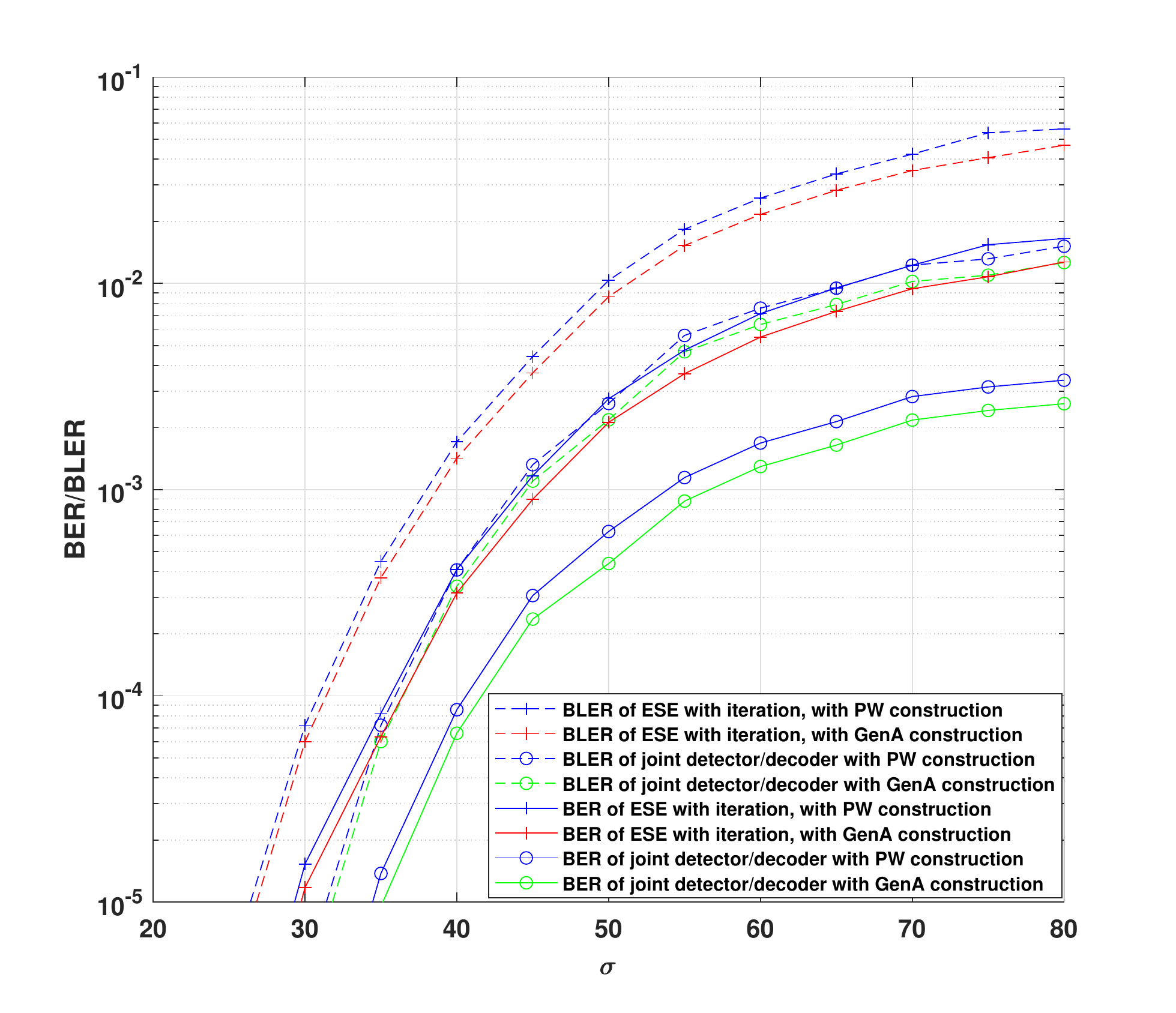}
  \caption{BER and BLER comparison of ReRAM system with the polar code, with different detectors and joint detection decoding schemes. The memory size is $M\times N=8\times8$, $p_{sf}=0.001$ and the code rate is 0.8.}\label{fig_001_8c}
\end{figure}
\begin{figure}
  \centering
  \includegraphics[width=3.2in, height=2.8in]{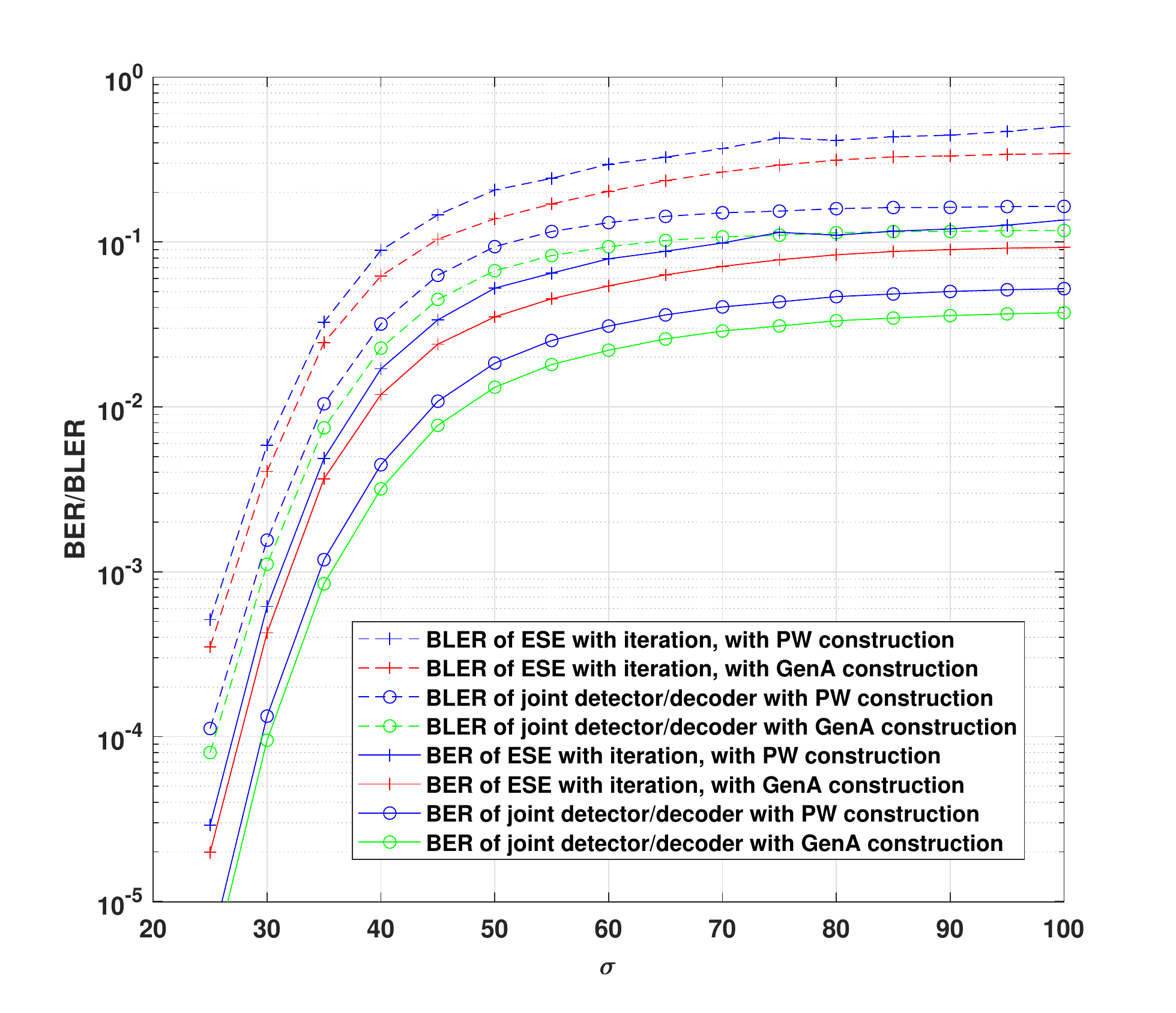}
  \caption{BER and BLER comparison of ReRAM system with the polar code, with different detectors and joint detection decoding schemes. The memory size is $M\times N=16\times16$, $p_{sf}=0.001$ and the code rate is 0.8.}\label{fig_001_16c}
\end{figure}

Next, we evaluate the performance of the ReRAM system with the proposed joint detection and decoding coding, with the polar code that is constructed using the GenA. The code rate is set to be $R_{ECC}=0.8$. \textcolor{black}{In the simulation results, the code construction is designed for a fixed channel noise value $\sigma =40$. In fact, what this value is has little effect on performance. Therefore, any fixed value can be selected in the working range. There are two reasons for this. Firstly, polar codes construction based on GenA is designed aimed at the asymmetric channel of ReRAM. The asymmetry of the ReRAM channel is caused by the sneak path, and noise has no effect on the symmetry. Secondly, Genetic algorithm is an algorithm in machine learning. According to the experience of the channel coding based on machine learning, the SNR has almost no effect on the learning effect during the training process. Therefore,  the code construction is designed for a fixed channel noise value.} The simulation results are shown in Fig. \ref{fig_001_8c} and Fig. \ref{fig_001_16c}. In the figures, we also include the error rate performance with the polar code with the same code rate, but is designed using the PW algorithm as a reference. Moreover, for cases with the ESE detector, for a fair comparison, we introduce the iteration of soft information between the BP decoder of the polar code with the ESE detector which is beyond the original work of \cite{17.code1}.  We name it ``ESE with iteration" in the figures. The number of iterations between the BP decoder and the BP/ESE detector is set to be 10.

It can be observed from the figures that the polar code constructed using the GenA that is tailored for the asymmetric ReRAM channel achieves better error rate performance than the conventional construction based on the PW algorithm. Moreover, the proposed BP based joint detection and decoding can achieve significantly better BER/BLER than the ESE detector which is with iteration of soft information with the BP decoder. In addition to the gain of the BP detector over the ESE detector, the message passing inside the joint graph of the BP detector and BP decoder improves the performance further, compared to the conventional soft iteration scheme where the detector ({\it i.e.} the ESE detector) is independent of the BP decoder.

\textcolor{black}{Furthermore, lognormal distributed resistance variations of the ReRAM channel is considered in this section. Experimental data shows that the resistance variation of ReRAM typically follows the lognormal distribution \cite{lognormal_1,lognormal_2,lognormal_3}.}

\begin{figure}
	\centering
	\includegraphics[width=3.2in, height=2.8in]{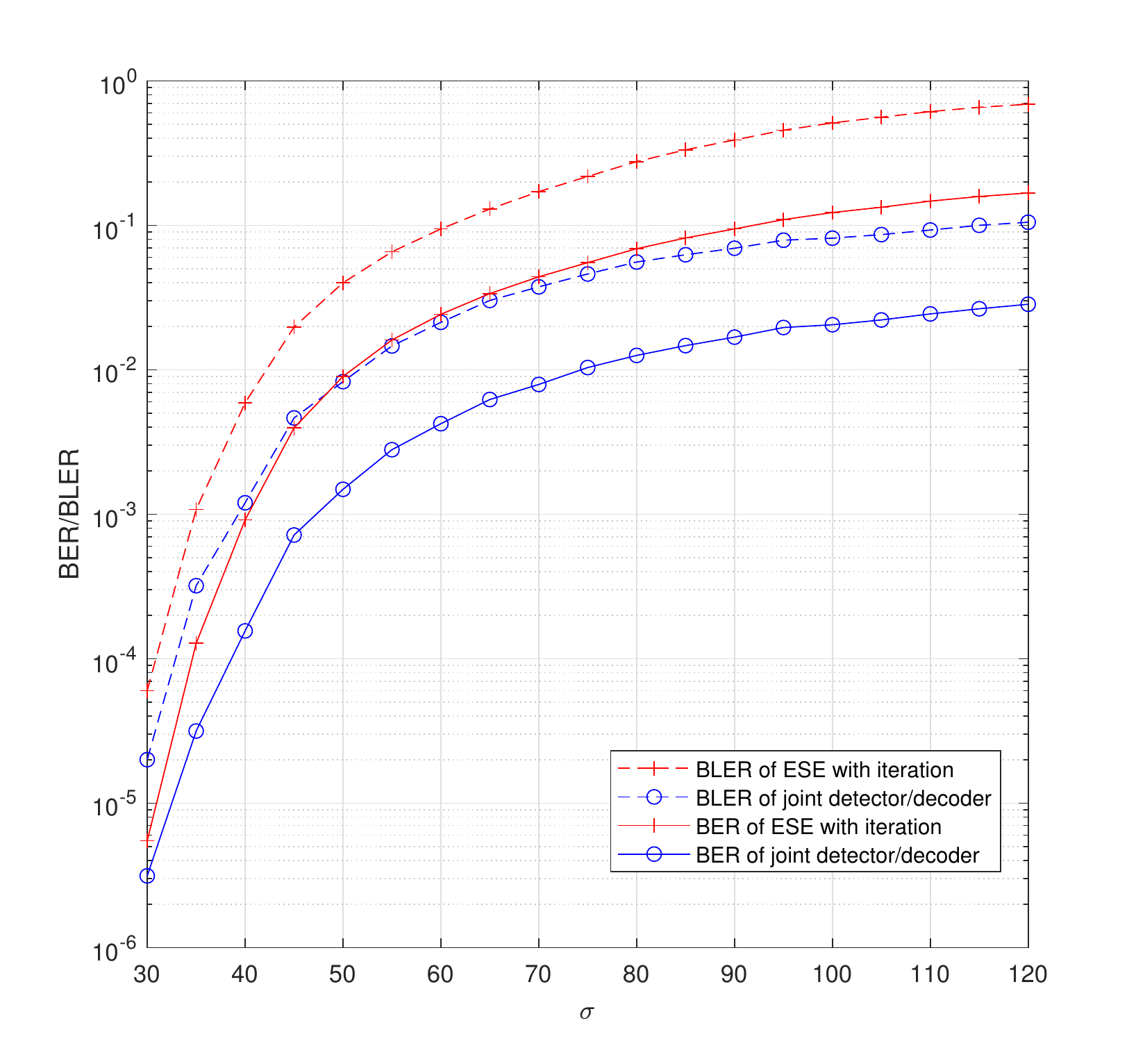}
	\caption{BER and BLER comparison of ReRAM system with the polar code, with different detectors and joint detection decoding schemes, where the resistance variation of ReRAM follows the lognormal distribution. The memory size is $M\times N=16\times16$, $p_{sf}=0.001$ and the code rate is 0.8.}\label{fig_001_16log}
\end{figure}

\textcolor{black}{The resulting signal ${y_{i,j}}$ follows the lognormal distribution, i.e., $\ln \left( {{y_{i,j}}} \right) \sim N\left( {{\mu _{i,j}},\sigma _{i,j}^2} \right)$ with}
\textcolor{black}{
\begin{equation}\label{log_u}
	{\mu _{i,j}} = \left\{ {\begin{array}{*{20}{l}}
			{{\mu _0} = \ln \left( {\frac{{R_0^2}}{{\sqrt {R_0^2 + {\sigma ^2}} }}} \right),}&{{r_{i,j}} = {R_0}}\\
			{\mu {'_0} = \ln \left( {\frac{{R{'_0}^2}}{{\sqrt {{R{'_0}^2} + {\sigma ^2}} }}} \right),}&{{r_{i,j}} = R{'_0}}\\
			{{\mu _1} = \ln \left( {\frac{{R_1^2}}{{\sqrt {R_1^2 + {\sigma ^2}} }}} \right),}&{{r_{i,j}} = {R_1}}
	\end{array}} \right.
\end{equation}}
\textcolor{black}{
	\begin{equation}\label{log_s}
		\sigma _{i,j}^2 = \left\{ {\begin{array}{*{20}{l}}
				{\sigma _0^2 = \ln \left( {1 + \frac{{{\sigma ^2}}}{{R_0^2}}} \right),}&{{r_{i,j}} = {R_0}}\\
				{\sigma {'_0}^2 = \ln \left( {1 + \frac{{{\sigma ^2}}}{{R{'_0}^2}}} \right),}&{{r_{i,j}} = R{'_0}}\\
				{\sigma _1^2 = \ln \left( {1 + \frac{{{\sigma ^2}}}{{R_1^2}}} \right),}&{{r_{i,j}} = {R_1}}
		\end{array}} \right.
\end{equation}}
\textcolor{black}{The PDFs of $y_{i,j}$ is then given as $f\left( {{y_{i,j}},{u_{i,j}},{\sigma _{i,j}}} \right) = \frac{1}{{{y_{i,j}}{\sigma _{i,j}}\sqrt {2\pi } }}{e^{ - \frac{{{{\left( {\ln {y_{i,j}} - {u_{i,j}}} \right)}^2}}}{{2\sigma _{i,j}^2}}}}$. By modifying the PDFs used in Section III.D, the BER/BLER performance of proposed scheme is simulated with lognormal distribution. As shown in Fig.\ref{fig_001_16log}, the proposed scheme has significant performance gain under the noise with lognormal distribution.}

\section{Conclusions}
In this paper, we have considered the sneak path interference, which is a major limiting factor of the ReRAM crossbar array. We have first proposed a novel BP based detector, whose Tanner graph is constructed based on the conditions for a sneak path to occur and the dependence of the states of the memory
cells that are involved in the sneak path. The graph of the designed BP detector has been further combined with that of the BP decoder of the polar codes to form a joint detection and decoding scheme. In the joint detection and decoding, the BP detector mainly mitigates the sneak path interference while the BP decoder corrects the errors caused by the channel additive noise. Effective polar codes are also constructed using the GenA, which is tailored for the joint detector and decoder over the ReRAM channel. Simulation results demonstrate that the BP detector can accurately detect the faulty cell selectors, and the proposed polar codes and the joint detector and decoder can achieve significantly better error rate performance than the prior art scheme.

\textcolor{black}{This work can be enriched from the following aspects. The implementation of the proposed BP detector remain largely open. We expect that the proposed scheme will reduce the implementation complexity  by reducing the number of nodes in graph. It is believed that this aspect deserves further investigation.}


\begin{thebibliography}{1}
\bibitem{1.ReRAM}
D. B. Strukov, G. S. Snider, D. R. Stewart, and R. S. Williams, ``The missing memristor found,'' Nature, vol. 453, no. 7191, p. 80, 2008.
\bibitem{2.Sneak_path}
M. A. Zidan, H. A. H. Fahmy, M. M. Hussain, and K. N. Salama, ``Memristor-based memory: The sneak paths problem and solutions,'' Microelectron. J., vol. 44, no. 2, pp. 176-183, 2013.
\bibitem{3.4}
M. A. Zidan, A. M. Eltawil, F. Kurdahi, H. A. Fahmy, and K. N. Salama, ''Memristor multiport readout: A closed-form solution for sneak paths,'' \textit{IEEE Trans. Nanotechnol.}, vol. 13, no. 2, pp. 274-282, Mar. 2014.
\bibitem{3.5}
X. Wang, M. Chen, Y. Shen, and X. Hu, ''A new crossbar architecture based on two serial memristors with threshold,'' \textit{2015 International Joint Conference on Neural Networks (IJCNN)}, Jul. 2015, pp. 1-6.
\bibitem{6.6}
P. O. Vontobel, W. Robinett, P. J. Kuekes, D. R. Stewart, J. Straznicky, and R. S. Williams, ''Writing to and reading from a nano-scale crossbar memory based on memristors,'' Nanotechnology, vol. 20, no. 42, Sep. 2009.
\bibitem{6.7}
S. Shin, K. Kim, and S.-M. Kang, ''Analysis of passive memristive devices array: Data-dependent statistical model and self-adaptable sense resistance for RRAMs,''\textit{ Proc. IEEE}, vol. 100, no. 6, pp. 2021-032, Jun. 2012.
\bibitem{6.8}
C. Liu and H. Li, ''A weighted sensing scheme for reram-based crosspoint memory array,'' \textit{Proc. IEEE Comput. Soc. Annu. Symp. VLSI (ISVLSI)}, Jul. 2014, pp. 65-70.
\bibitem{6.9}
M. Zidan et al., ''Single-readout high-density memristor crossbar,'' \textit{Sci. Rep.}, vol. 6, Jan. 2016, Art. no. 18863.
\bibitem{10.theo}
P. P. Sotiriadis, ''Information capacity of nanowire crossbar switching networks,'' \textit{IEEE Trans. Inf. Theory}, vol. 52, no. 7, pp. 3019-3032, Jul. 2006.
\bibitem{11.theo_ReRAM}
Y. Cassuto, S. Kvatinsky, and E. Yaakobi, ''Information-theoretic sneak path mitigation in memristor crossbar arrays,'' \textit{IEEE Trans. Inf. Theory}, vol. 62, no. 9, pp. 4801-813, Sep. 2016.
\bibitem{12.faulty_selector}
Y. Deng et al., ''RRAM Crossbar Array With Cell Selection Device: A Device and Circuit Interaction Study,'' \textit{IEEE Trans. Electron Devices.}, vol. 60, no. 2, pp. 719-726, Feb, 2013.
\bibitem{13.sf1}
Y. Ben-Hur and Y. Cassuto, ''Detection and coding schemes for sneakpath interference in resistive memory arrays,'' \textit{IEEE Trans. Commun.}, vol. 67, no. 6, pp. 3821-3833, Feb. 2019.
\bibitem{14.sf2}
Z. Chen, C. Schoeny, and L. Dolecek, ''Pilot assisted adaptive thresholding for sneak-path mitigation in resistive memories with failed selection devices,'' \textit{IEEE Trans. Commun.}, vol. 68, no. 1, pp. 66-81, Jan. 2020.
\bibitem{15.noise1}
A. Chen and M.-R. Lin, ''Variability of resistive switching memories and its impact on crossbar array performance,'' \textit{Proc. Int. Rel. Phys. Symp., Monterey}, CA, USA, Apr. 2011.
\bibitem{16.noise2}
Y. Ben-Hur and Y. Cassuto, ''Detection and coding schemes for parallel interference in resistive memories,'' \textit{Proc. IEEE Int. Conf. Commun. (ICC)}, Paris, France, May 2017, pp. 1.
\bibitem{17.code1}
G. Song, K. Cai, X. Zhong, J. Cheng, ''Performance Limit and Code Design for Resistive Random-Access Memory Channels'', \textit{arXiv}:2005.02601, May. 2020.  Accepted by \textit{IEEE Trans. Commun.}, Jan. 2021.
\bibitem{18.code2}
M. Zorgui, M. E. Fouda, Z. Wang, A. M. Eltawil and F. Kurdahi, ''Non-Stationary Polar Codes for Resistive Memories,'' \textit{2019 IEEE Global Communications Conference (GLOBECOM)}, Waikoloa, HI, USA, 2019, pp. 1-6
\bibitem{19.LDPC}
D. J. C. MacKay, ''Good error-correcting codes based on very sparse matrices'', \textit{IEEE Trans. Inf. Theory}, vol. 45, no. 3, pp. 399-431, Mar. 1999.
\bibitem{20.Polar}
E. Arikan, ''Channel polarization: A method for constructing capacity-achieving codes for symmetric binary-Input memoryless channels,'' \textit{IEEE Trans. Inform. Theory.}, vol. 55, no. 7, pp. 3051-3073, Jul. 2009.
\bibitem{6.BP}
J. Pearl,  \textit{Probabilistic reasoning in intelligent systems: Networks of
plausible inference}, San Mateo, CA, USA: Morgan Kaufmann, 1987.
\bibitem{21.GenA}
Huang, Lingchen , et al. ''AI Coding: Learning to Construct Error Correction Codes.'' \textit{IEEE Trans. Comm} vol. 68, no. 1, Jan. 2020.
\textcolor{black}{\bibitem{lognormal_1}
A. Fantini et al., ''Intrinsic switching variability in HfO2 RRAM.'' \textit{IEEE International Memory Workshop}, 2013, pp. 30–33.}
\textcolor{black}{\bibitem{lognormal_2}
S. Yu, X. Guan, and H.-S. P. Wong, ''On the Switching Parameter Variation of Metal Oxide RRAM—Part II: Model Corroboration and Device Design Strategy,'' \textit{IEEE Trans. Electron Devices}, vol. 59, no. 4, pp. 1183–1188, 2012.}
\textcolor{black}{\bibitem{lognormal_3}
P. Li, G. Song, K. Cai and Q. Yu, ''Across-Array Coding for Resistive Memories with Sneak-Path Interference and Lognormal Distributed Resistance Variations,'' \textit{IEEE Commun. Lett},. Early Access.}
\end{thebibliography}
\end{document}